\documentclass[12pt,a4paper]{article}
\usepackage[left=3cm, right=3cm, top=3cm, bottom=3.5cm]{geometry}
\usepackage[utf8]{inputenc}
\usepackage{graphicx}
\usepackage{subcaption}
\usepackage{overpic}
\usepackage{amsmath}
\usepackage{amssymb}
\usepackage{placeins}

\usepackage{color} 
\usepackage{placeins} 
\usepackage{hyperref}

\title{Vacuum stability of Froggatt-Nielsen models in the EFT regime}
\author{}

\newcommand{\lag}{\mathcal{L}}
\newcommand{\Higgs}{H}
\newcommand{\Flavon}{S}
\newcommand{\order}[1]{\mathcal{O}\left(#1\right)}
\newcommand{\orderone}{\order{1}}

\newcommand{\quarkL}[2]{#1_L^{#2}}
\newcommand{\barquarkL}[2]{\bar#1_L^{#2}}
\newcommand{\quarkR}[2]{#1_R^{#2}}
\newcommand{\barquarkR}[2]{\bar#1_R^{#2}}

\newcommand{\be}{\begin{equation}}
\newcommand{\ee}{\end{equation}}
\newcommand{\bea}{\begin{eqnarray}}
\newcommand{\eea}{\end{eqnarray}}
\newcommand{\nn}{\nonumber}

\newcommand{\lp}{\left(}
\newcommand{\rp}{\right)}

\begin{document}

\begin{titlepage}
\begin{flushright}
DESY 19-158\\
\end{flushright}
\vspace{.3in}

\vspace{1cm}
\begin{center}
{\Large\bf\color{black}
Vacuum stability of Froggatt-Nielsen models
}\\
\bigskip\color{black}
\vspace{1cm}{
{\large F.~Giese, T.~Konstandin}
\vspace{0.3cm}
} \\[7mm]
{\it 
{DESY, Notkestra\ss e 85, D-22607 Hamburg, Germany} \\
}
\end{center}
\bigskip

\vspace{.4cm}

\begin{abstract}
We discuss vacuum stability in Froggatt-Nielsen (FN) models. One concern in FN models is that for large flavon VEVs the running of the quartic Higgs coupling is enhanced what might lead to a more severe instability compared to the Standard Model (SM). We study this issue using the renormalization-group improved scalar potential. Another issue is that the mixing between the Higgs and the flavon can potentially destabilize the potential. However, taking current bounds on the flavon phenomenology into account, we find that both effects do not lead to an instability that is more severe than in the SM. 
\end{abstract}

\bigskip

\end{titlepage}

\newpage
\tableofcontents
\newpage

\section{Introduction}

One of the open questions in the Standard Model (SM) of particle physics stems from the hierarchy of the masses and couplings in the fermionic sector. The fermion masses are generated through a Yukawa coupling to the Higgs field that obtains a vacuum expectation value (VEV) and breaks the electroweak symmetry spontaneously. The large range of fermion masses - from the 512 keV of the electron to the 175 GeV of the top quark - then requires a large range of six orders of magnitude of Yukawa couplings.

One way to explain this large hierarchy of couplings is the Froggatt-Nielsen mechanism~\cite{Froggatt:1978nt}. The central idea is to assign different charges of a new $U(1)_{FN}$ symmetry group to the different particles of the SM. When the left- and right-handed fields are assigned different charges, the Yukawa couplings of the SM can then only be produced by breaking the new $U(1)_{FN}$ symmetry. For example, the symmetry can be spontaneously broken by giving a VEV to a flavon field $S$ that is also charged under $U(1)_{FN}$. The Yukawa couplings of the SM are then generated by higher dimensional operators with certain powers of $\epsilon \sim \left< S \right>/M$ where $M$ is the typical mass scale of the flavon sector that produces the Yukawa couplings. If this ratio is somewhat smaller than unity, the large hierarchies in the Yukawa couplings can be explained with not too baroque charge assignments~\cite{Feruglio:2015jfa,Babu:2009fd}.

This construction has several appealing features. Besides the hierarchies in the Yukawa couplings, one main motivation for this construction are the patterns in the CKM matrix. Naturally, the FN mechanism predicts the Cabibbo angle from the mass hierarchies in the first two generations ($\theta_{12}\sim \epsilon$) and also the smallness of the mixing angles involving the third generation ($\theta_{13}\sim \epsilon^2$, $\theta_{23}\sim \epsilon^3$). 

Moreover, the dynamic generation of the Yukawa couplings can be potentially utilized in baryogenesis~\cite{Baldes:2016rqn, Baldes:2016gaf, Bruggisser:2017lhc}. For example, one possibility is that the spontaneous breaking of the $U(1)_{FN}$ symmetry is intertwined with electroweak symmetry breaking~\cite{Baldes:2016gaf}. This 
can strengthen the electroweak phase transition and also lead to new CP-violating sources that drive baryogenesis~\cite{Bruggisser:2017lhc}.  

The main motivation of the current work is to study implications of the FN setup for vacuum stability. Naively, if the Yukawa couplings are considered to be dependent on the flavon VEV, $y(\left< S \right>/M)$, all Yukawa couplings will be of order unity for large flavon VEVs, $\left< S \right> \sim M$. This potentially has disastrous effects. For once, large Yukawa couplings will drive the Higgs quartic coupling $\lambda_h$ to negative values very quickly and the electroweak vacuum becomes unstable. 
In particular, in the context of models with varying Yukawa couplings, where the UV theory is not known, the stability of the effective potential is an open question~\cite{Braconi:2018gxo}.

We will see that stability is actually not an issue in FN models for several reasons.  The first reason is that also the flavon field will contribute to the scalar potential. Displacing the VEV from its minimum $\left< S \right> \sim \epsilon M$ to $\left< S \right> \sim M$ will increase the scalar potential. How large this effect is depends on the flavon mass. Second, also the Higgs quartic coupling can become a function of the flavon VEV, $\lambda(\left< S \right>/M)$. This effect will counteract the increased running of the quartic coupling in case $\left< S \right>/M$ is of order unity. 

In order to study this issue, we will use the renormalization group (RG) improved effective potential. This will serve to demonstrate above points but not be competitive with current studies of vacuum stability in the SM that use the $\beta$ functions up to three loop order \cite{Degrassi:2012ry, Bednyakov:2015sca}. In the current study, we have the additional complication that we have to deal with two scalar VEVs. There are several proposals in the literature how to obtain a RG-improved effective potential in this circumstance~\cite{Einhorn:1984ej, Bando:1992wy, Ford:1996yc, Casas:1998cf, Chataignier:2018aud}. We will see that we do not require these techniques here and discuss in detail how to resum the leading logs in our context. 


\section{The Froggatt-Nielsen model\label{sec:FNmodel}}
\label{sec:FNEffectivePotential} 
As a model we consider an extension of the SM where the Yukawa coupling of the bottom quark is generated by the flavon sector while the top quark sector is as in the Standard Model.
The Lagrangian in the scalar sector contains the terms
\be
\lag\ =-m_h^2 \, H^\dagger H+ \lambda_h \left(H^\dagger H\right)^2- m^2_s S^* S
+ \lambda_s \left(S^*S\right)^2+ \lambda_m  H^\dagger H  \, S^* S  \, ,
\ee
which gives rise to the tree level scalar potential for the real degrees of freedom
\be
\label{eq:V0}
V_0(\phi, s) = -\frac{m_h^2}{2} h^2 + \frac{\lambda_h}{4} h^4
-\frac{m_s^2}{2} s^2 + \frac{\lambda_s}{4} s^4
+ \frac{\lambda_m}4 s^2 h^2 \, . 
\ee

The complex flavon field $S$ also contains a pseudo-scalar degree of freedom $a$ that would be massless after the spontaneous breaking of the (global) $U(1)_{FN}$ symmetry. We will neglect this particle in our analysis and assume that it 
obtains a soft mass term through an explicitly $U(1)_{FN}$-breaking term that is above all relevant scales that we aim to study. In a more realistic set-up the pseudo-scalar would be expected to be very light and would have phenomenological consequences \cite{Bauer:2016rxs}, which we also discuss in appendix \ref{sec:ExperimentalBounds}.

The top quark mass is induced through a usual Yukawa term\footnote{Throughout the paper we use a slightly lighter top quark mass $y_t=0.95$ instead of $y_t=1$. We do this in order to recover the known instability scale from three-loop calculations that lead to additional threshold effects not present at 1-loop.}%
\be
m_t^2  = \frac{y^2_t}{2} h^2 \, ,
\ee
while the mass of the bottom quark is generated in the flavon sector. We consider a charge assignment where the bottom quark mass requires two insertions of the flavon VEV and two corresponding (heavy) quarks $F$. The Lagrangian contains the term 
\be
\begin{pmatrix}
\bar q_L^{(0)} \\
\bar F^{(0)}_L \\
\bar F^{(-1)}_L
\end{pmatrix}^T
\begin{pmatrix}
0 & Y_b\,  H^{(0)} & 0\\
0 & M/\sqrt{2} & Y_B\, S^{(-1)} \\
Y_B\, S^{(-1)} & 0 & M/\sqrt{2}
\end{pmatrix}
\begin{pmatrix}
b_R^{(2)}\\
F^{(0)}_R\\
F^{(1)}_R
\end{pmatrix},
\ee
where we indicated the particles FN charges in the superscripts. The corresponding mass matrix reads
\be
M_Q^\dagger M_Q=\frac{1}{2}\begin{pmatrix}
Y_b^2h^2 &   Y_bh M & 0\\
 Y_bh M &  M^2+Y_B^2 \, s^2 &  Y_B\ sM\\
0 &  Y_B \,s M & M^2+Y_B^2 \,s^2
\end{pmatrix} \, .
\ee
For $\phi \ll s$ and $\left(Y_B\ s/M\right) \equiv \epsilon$ this matrix has two eigenvalues of order $M^2$ and one light eigenvalue that constitutes the bottom quark, 
\be
m_b^2 \simeq \frac{Y_b^2}2 h^2 \epsilon^4 \, .
\ee
Realistic bottom masses can be obtained for $\epsilon \sim 0.2$ and $Y_b \sim 1$. For convenience, we set $Y_B=1$ in the following. 
As a first assessment of the vacuum stability of the model we will calculate the one-loop effective potential (in Landau gauge). The relevant degrees of freedom besides the quark sector are the $W$, $Z$ gauge bosons, with masses 
\be
m_W^2 = \frac14 g^2 h^2, \quad
m_Z^2 = \frac14 (g^2 + g^{\prime2}) h^2 .
\ee
Furthermore, the scalar sector contains the Higgs boson $h$, the flavon $s$ and Goldstone bosons $\chi$
with the masses
\be
\label{eq:m2_scalars}
m^2 = 
\begin{pmatrix}
3 \lambda_h h^2 - m_h^2 + \lambda_m s^2/2 & \lambda_m h s\\
 \lambda_m h s & 3 \lambda_s s^2 - m_s^2 + \lambda_m h^2 /2\\
\end{pmatrix} \, ,
\ee
and 
\be
m_\chi^2 = \lambda_h h^2 - m^2 + \lambda_m s^2/2 \, .
\ee
In principle, the Goldstones would mix with the pseudo-scalar residing in the flavon field. As explained before, we 
decoupled this field by giving it an explicit mass term.\newline
 Alternatively, one can eliminate the mass parameters and write these mass matrices in terms of the VEVs in the minimum of the potential. We denote the two eigenvalues of (\ref{eq:m2_scalars}) as $m_\phi$ and $m_{\sigma}$ according to the states that are predominantly Higgs and flavon.

Minimizing the potential in (\ref{eq:V0}) leads then to 
\be
m^2 = 
\begin{pmatrix}
\lambda_h (3 h^2 - v^2) + \lambda_m (s^2 - w^2)/2 & \lambda_m h s\\
 \lambda_m h s & \lambda_s (3 s^2 - w^2) + \lambda_m (h^2 - v^2)/2 \\
\end{pmatrix} \, ,
\ee
and 
\be
m_\chi^2 = \lambda_h (h^2 - v^2) \, ,
\ee
where $v$ and $w$ denote the minima of the VEVs $h$ and $s$.

For the one-loop effective potential, we use the known contributions from Coleman and Weinberg~\cite{Coleman:1973jx} in $\overline{MS}$ regularization, namely ($\kappa=1/(16\pi^2)$)
\be
V_1=\sum_i \frac{\kappa g_i}{4} m_i^4 \left( \log  [m_i^2/\mu^2] - c_i \right) \, .\label{eqV1correction}
\ee
Here, $g_i$ denotes the number of degrees of freedom ($-12$ for top, $b$ and $F$ quarks, $3$ and $6$ for $W$- and $Z$-bosons, $3$ for Goldstone bosons, $1$ for the remaining scalars). The $c_i$ are constants that are however irrelevant since we renormalize such that the VEVs in the minima are held constant. As parameters we hence use the two VEVs -- $v$ and $w$ -- the physical masses of the Higgs, $m_\phi$, and the flavon $m_\sigma$ and their mixing $\theta$. The mixing angle is related to the couplings and VEVs via
\be
\tan 2\theta=\frac{\lambda_{m}v w}{\lambda_h v^2-\lambda_s w^2} \, .
\ee 
Note that in this convention for $v\ll w$ negative $\lambda_m$ leads to a positive mixing angle. Further in our analysis we restrict ourselves to the case of the Higgs being the lighter mass eigenvalue $\theta\in\left(-\frac{\pi}{4},\frac{\pi}{4}\right)$.\\

\begin{figure}[t]
	\begin{subfigure}[t]{0.31\textheight}
		\includegraphics[width=0.9\textwidth]{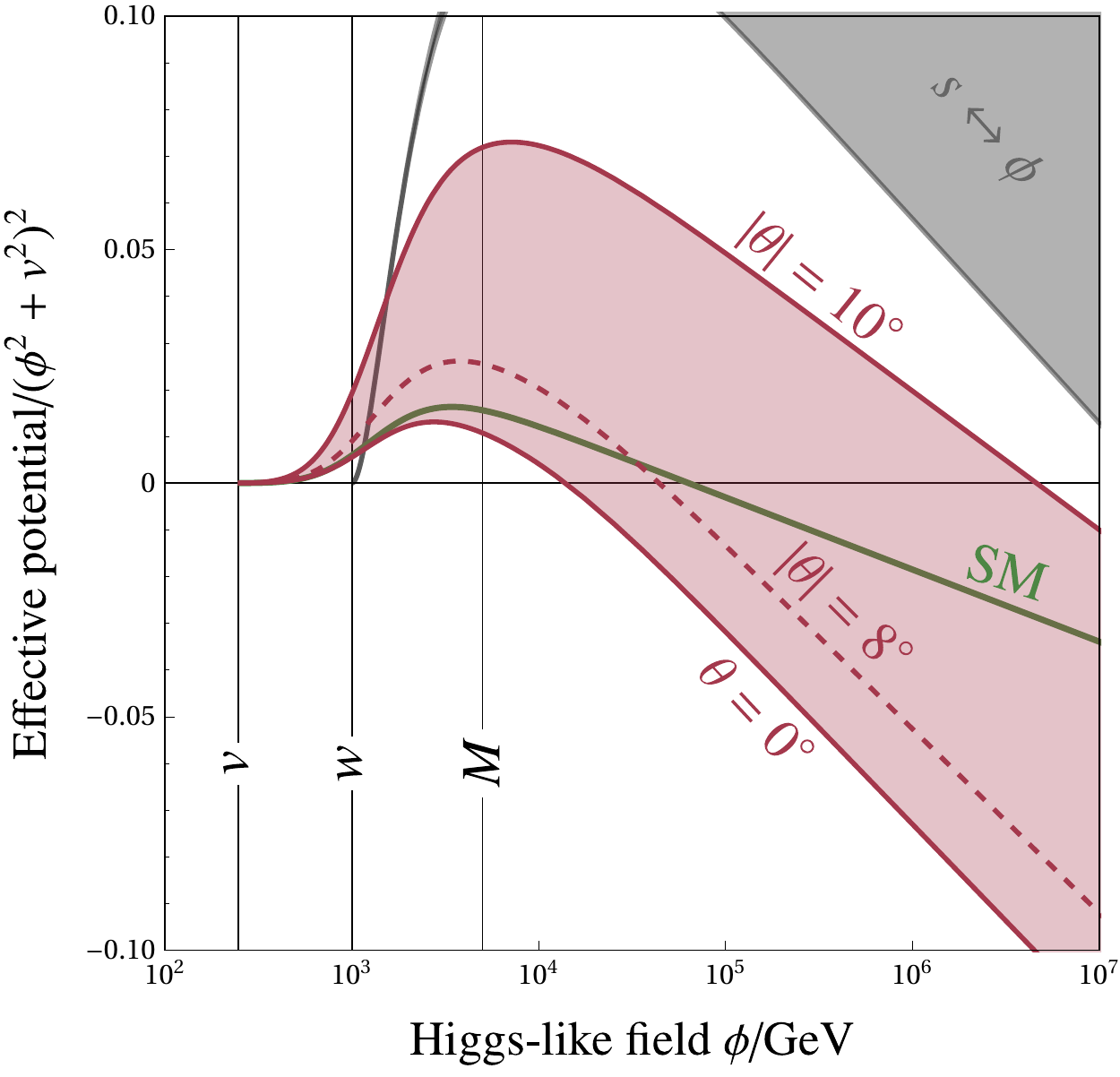}
	\end{subfigure}
	\begin{subfigure}[t]{0.31\textheight}
		\includegraphics[width=0.9\textwidth]{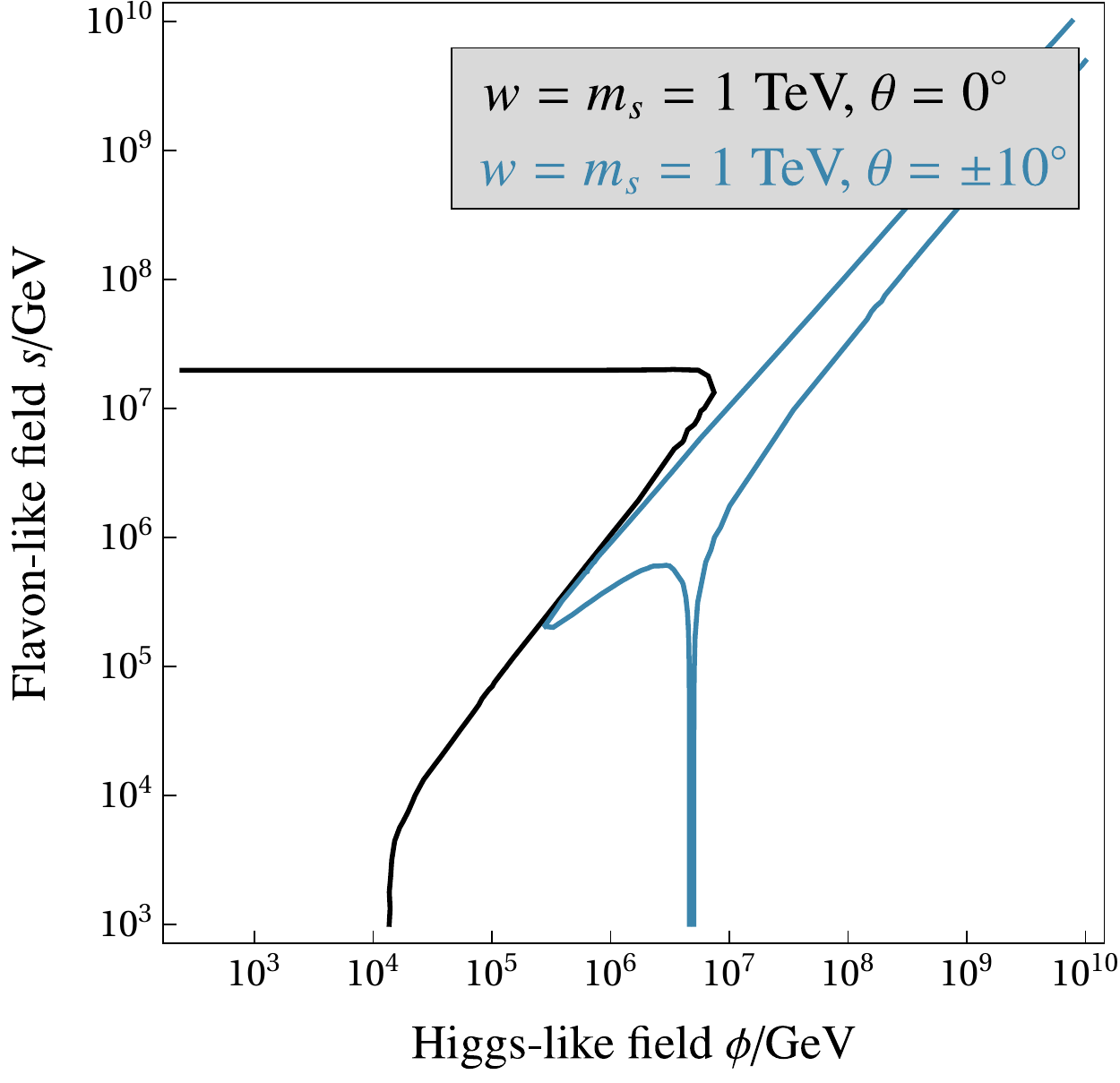}
	\end{subfigure}
	\caption{\small
The left panel shows the comparison between the effective potential of the SM and the model invoking a Froggatt-Nielsen mechanism for the $b$ quark in $h$-direction. The flavon sector is fixed as $m_\sigma=w=1\text{ TeV}$ whereas the scalar mixing angle $\theta_s$ varies (red band). The Froggatt-Nielsen scale is $M={5}\text{ TeV}$. The gray curve corresponds to the potential in $s$-direction. The right panel shows the zero-potential lines in the model with and without mixing.}
	\label{figureEffectivePotential1}
\end{figure}
Once the effective potential is known, the instability scale $\Lambda_I$ is read off from the scale where the effective potential hits zero~\footnote{The VEV at which the potential vanishes is actually not an observable and suffers from gauge-dependence. However it turns out that in Landau gauge this scale is very close to the mass scale where new degrees of freedom must be introduced to (absolutely) stabilize the potential~\cite{Espinosa:2016uaw, Espinosa:2016nld}.}. The left panel of figure~\ref{figureEffectivePotential1} shows the comparison between the SM vacuum instability scale and a low energy realization~($M=5\text{ TeV}$) of the Froggatt-Nielsen mechanism near meson mixing limits~(\ref{sec:BoundsFromMesonMixing}). The TeV value chosen for the FN scale M keeps the error by the inclusion of heavy degrees of freedom below the UV scale to a minimum (we will construct an RG-improved effective potential which can handle large UV scales $M$ in the next section). The red curve shows the potential in $h$-direction whereas the gray line shows the potential in $s$-direction which is not present in the SM. The case~$\theta=0^\circ$ offers the most severe instability. It lies in direction of the $h^4$ operator and thus can be directly compared with the SM instability $\Lambda_I^{\text{SM}}$ (green line). The FN instability (for $\theta=0^\circ$) is lower because of additional log contributions from the heavy quarks. Still, for the chosen UV completion the fermion logarithm can be easily counteracted by introducing mixing in the scalar sector. Thus there is the possibility that the stability issue in the SM might not even be present in a UV completed FN setup. 

The right panel of figure~\ref{figureEffectivePotential1} shows the stabilizing effect from increased mixing in the $h-s$ plane such that also the effect from the $h^2 s^2$ operator becomes visible. The shown lines correspond to the respective instability contours of vanishing mixing (black) and non-vanishing mixing (blue). Two things can be observed. The first effect is a feature on the diagonal which is due to the non-vanishing tree level operator $h^2 s^2$ whose coefficient is fixed to be $\lambda_m$ in our notation. For $\lambda_m\neq0$ the operator can be stabilizing or destabilizing depending on the sign of $\lambda_m$. The second effect is the stabilizing effect caused by mixing of the flavor eigenstates which does not depend on the sign of $\lambda_m$. This effect can overturn the negative contribution from the extended fermion sector in $h$-direction. In general the barrier in $s$-direction is high and large in case of $m_s\sim M\gg m_h$. This reduces the tunneling amplitude calculation effectively to a 1D path problem in $h$-direction (with a fixed flavon VEV). Thus to create an instability in the EFT region below the heavy quark threshold one can use two ingredients, a negative $\lambda_m$ and a small flavon mass in comparison to $M$. However, once $M$ and $m_\sigma$ are decoupled, one should do proper RG-running between the scales.

\FloatBarrier
\section{Effective field theory and renormalization group running}
\label{sectionEffectiveFieldTheoryAndRG}

We have seen in the last section that the FN extension in principle does not seem to have an issue with vacuum stability that is more serious than in the SM. In this section, we will provide a renormalization group improved analysis of the effective potential. There are two motivations for this. First, the RG-improved potential deviates significantly from the one-loop analysis -- even in the SM. This is due to the fact that the main running of the quartic coupling comes from the Yukawa coupling of the top and enters in its fourth power. Since the Yukawa coupling themselves run, a small change in the top Yukawa coupling can have a large impact on the potential. For example the potential turns negative around $h \sim 10^4$ GeV in the one-loop approximation while this happens at $h \sim 10^8$ GeV in the RG-improved analysis. The second reason to present the RG-improved analysis is that the effective field theory (EFT) being setup in the RG framework elucidates why the FN mechanism has minimal impact on vacuum stability below the FN threshold.

The basic observation of the RG analysis is that the observables that are derived from the effective action (in gradient expansion),
\be
S \simeq \int d^4x \, \frac{Z_h}2 \partial_\mu h \partial^\mu h
+ \frac{Z_s}2 \partial_\mu s \partial^\mu s - V(s,h) \, , 
\ee
should be independent from the renormalization scale $\mu$ that explicitly appears in the loop corrections, as in equation (\ref{eqV1correction}). 
This explicit dependence hence can be absorbed into the parameters of the theory: the couplings, mass terms, cosmological constant and wave function normalizations $Z$. To simplify matters, one can rescale the fields such that the kinetic terms are always canonically normalized. The effective potential has then to be scale-independent by itself and the scale dependence in the wave function normalizations show up as the anomalous dimensions in the RG equation of the effective potential
\be
0 = \mu\frac{d}{d\mu} V = 
\left( \mu\frac{\partial}{\partial\mu} 
+ \sum \beta_\lambda \frac{\partial}{\partial \lambda}
+ \gamma_h \frac{\partial}{ \partial h}
+ \gamma_s \frac{\partial}{ \partial s}
\right) V \, ,
\ee
where we denote the couplings and mass parameters of the theory collectively as $\lambda$, $\beta = \mu \frac{\partial}{\partial \mu} \lambda$ and $\gamma_h = \mu\frac{\partial}{\partial \mu} Z_h$ and likewise for the field $s$. 
This relation can in turn also be used to deduce some of the $\beta$-functions. This is due to the fact that the leading explicit $\mu$-dependence results from the loop corrections while the other terms in the RG equation obtain contributions from the tree level potential. We will use this to deduce the $\beta$-functions of the scalar operators in the EFTs that we will set up in the following. A complete list of beta functions can be found in appendix \ref{appendixBetaFunctions}.

\subsection{Hierarchy of effective field theories}

\begin{figure}[t]
	\centering
	\includegraphics[width=0.95\textwidth]{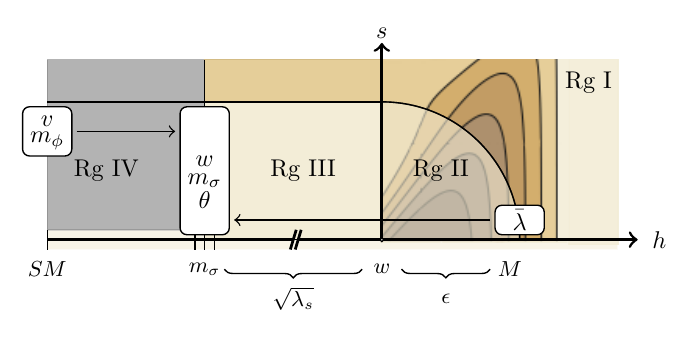}
	\caption{
\small Schematic construction of the RG-improved effective potential in case of large scale separation $v \ll m_\sigma\ll w$ as described in the text. Heavy fermions that are integrated out at $\mu_{UV}\sim M$ lead to higher dimensional operators e.g. $\bar\lambda\ h^4 s^8/M^8$  (which is the operator of the same dimension as the $b$ quark mass term). $\bar\lambda$ also runs, mainly due to the one-loop contribution by the bottom quarks, which we take into account. For scales below $m_\sigma$ the new physics direction is ignored, and the resummation of logs can be performed the same way as in the SM. We define the free parameters of the model at the scales shown in the picture. The scale separation between the different mass scales in the FN mechanism is set by $\sqrt{\lambda_s}$ and $\epsilon$. The different regions in the effective potential landscape are described in the text.} 
	\label{figureEFTconstruction}
\end{figure}

We start the RG analysis at the UV scale and then successively lower the scale and construct the relevant EFTs on the way.
In the present context there are two VEVs ($h$ and $s$) which complicates the analysis. The spectrum depends on the VEVs and this has an impact on how the EFT has to be set up. In general, there is no straight-forward way to resum all the leading logs using the usual RG methods (not even in the SM, but large logs from light particles are not expected to contribute strongly to the potential, since this would lead to IR divergences in the limit of vanishing masses). Several papers in the literature have proposed solutions to this problem that rely on a generalized $\overline{\text{MS}}$ multiscale approach~\cite{Einhorn:1984ej,Bando:1992wy,Ford:1996yc,Casas:1998cf,Chataignier:2018aud}. We will follow a related but slightly different route here. Motivated by the results in the one-loop approximation, one can study the vacuum decay only in the two directions where one of the two VEVs is constant, namely along $(h,w)$ and $(w, s)$. This allows us to construct the EFTs in all relevant regions of the scalar field space as we will see below. A schematic representation of our procedure and its different regions can be found in figure~\ref{figureEFTconstruction}. 

\paragraph{The UV (Region I).} Consider the UV scale, where $h$ is of similar size as $M$ (region I). In this regime, almost all masses are of order $M$, the sole exception being the flavon that could potentially (in case of very small coupling) be significantly lighter than that. However, the resulting large logs are not so relevant. First, parametrically the singlet mass scales as $m_\sigma^2 \sim \lambda_s s^2$ which is typically not so far from $M$. Besides, when the flavon is much lighter than $M$, its overall contribution is small and hence not really relevant. In summary, the fact that the logs in this sector are not properly resummed cannot have a large impact on the total potential.

The contribution from the FN quark sector to the running of the different operators can be read of from 
\be
\label{eq:UV_running_FNq}
\mathrm{Tr}\ (M_Q^\dagger M_Q)^2 = \frac{1}{4}\left(2 M^2 s^2 + 2 (M^2 + s^2)^2 + 2 M^2 Y_b^2 h^2 +  Y_b^4 h^4\right) \, .
\ee
Hence in this regime, the FN quarks contribute significantly to the running of the cosmological constant (CC), the Higgs and flavon quartic couplings as well as the Higgs mass sector (which will induce a hierarchy problem). Accordingly, if all SM fermions are implemented in the FN sector, a very large negative contribution to the Higgs quartic will quickly drive the effective potential to negative values which we have seen before in figure~\ref{figureEffectivePotential1}. However one should be  cautious at this point. The exact result relies on the concrete FN realization we have chosen. In principle, the FN sector can include a large number of bosonic degrees of freedom beyond the threshold $M$ with sizable couplings to the Higgs that can counteract this effect. This number probably does not even need to be large in comparison to the $\order{>60}$ heavy quarks induced in a full FN treatment of the SM as has been argued before. Since any conclusions concerning  the UV are so strongly model-dependent we will ignore any instability that arises beyond the FN scale $M$ even though they are severe in the sense that the $s$ direction is not stabilized by the SM particle content. Our attention will instead be on the question whether the light FN $b$ quark induces an instability at scales below $M$.\\

\paragraph{The EFT below the FN quarks (Regions II+III).}
Below $M$ the heavy quarks are integrated out and we do the corresponding matching to an effective field theory (called FN EFT in the following). The matching scale is of order $\mu\sim w\sim M/5$. In principle the separation between $w$ and $M$ could be larger increasing the size of region II in figure \ref{figureEFTconstruction} which would on the other hand require $\order{>1}$ Yukawa couplings and/or a different charge assignment in the FN sector. 
In order to perform the matching, the one-loop contributions from the FN quarks can be expanded in $h/M$ and $s/M$. This expansion will lead to a set of operators of form 
\be
12 \kappa \lambda_{(m,n)} \frac{(Y_bh)^m s^n}{M^{m+n-4}} \, .
\ee
We enumerate all operators up to order $O(h^4)$ and $O(s^8)$:
\bea
\label{eq:heavy_stuff}
&\lambda_{(0,0)} = \frac1{16}(-1+ 2L_M) \, , \quad
\lambda_{(0,2)} = \frac1{8} (1+ 3L_M)  \, , \quad& \nn \\
&\lambda_{(0,4)} = \frac{23}{96} + \frac18 L_M  \, , \quad
\lambda_{(0,6)} = -\frac{1}{480}  \, , \quad
\lambda_{(0,8)} = \frac{17}{2240}  \, , \quad& \nn \\
&\lambda_{(2,0)} = \frac1{8} L_M \, , \quad
\lambda_{(2,2)} = \frac1{16}   \, , \quad& \nn \\
&\lambda_{(2,4)} = \frac{1}{32}  \, , \quad
\lambda_{(2,6)} = -\frac{1}{24}  \, , \quad
\lambda_{(2,8)} = \frac{1}{64}  \, , \quad& \nn \\
&\lambda_{(4,0)} = \frac1{16}(1+ L_M) \, , \quad
\lambda_{(4,2)} = -\frac1{48}   \, , \quad& \nn \\
&\lambda_{(4,4)} = -\frac{29}{960}  \, , \quad
\lambda_{(4,6)} = \frac{43}{3360}  \, , \quad
\lambda_{(4,8)} = \frac{1}{40320} (7067 - 2520 L_M)  \, , \quad& 
\eea
where we defined $L_M = \log(M^2/2\mu^2)$. Furthermore, there is the contribution from the bottom
\be
\label{eq:contr_bottom}
12\kappa \frac{Y_b^4 h^4 s^8}{16 M^8} 
\left( \log \left[ \frac{Y_b^2h^2s^4}{2\mu^2M^4} \right] - \frac32 \right) \, .
\ee
Notice that the logarithmic contribution reproduces explicitly the $\beta$-function already inferred in (\ref{eq:UV_running_FNq}). The only non-renormalizable operators that contain logarithmic contributions are $\lambda_{(4,8)}$ and the bottom contribution in (\ref{eq:contr_bottom}). As required the $\mu$-dependence cancels between these two contributions since the UV theory is renormalizable and the logs are related to the divergences in dimensional regularization.
So logarithmic contributions can only appear in renormalizable operators.

This changes in the EFT that is not renormalizable. The EFT at one-loop level will contain the contribution from the bottom quark (\ref{eq:contr_bottom}) but not the logs produced by the heavy quarks in (\ref{eq:heavy_stuff}). This is consistent with the fact that the $\beta$-function in the EFT will not contain the contributions from the heavy quarks. At the same time, the EFT will have to include the higher dimensional operators from (\ref{eq:heavy_stuff}) in order to make the matching possible. 
These operators contribute to the Higgs and flavon masses and hence also will participate in the one-loop running and also run themselves. For example, the flavon one-loop contribution will contribute a term of form $\lambda_{(2,6)}\lambda_{(0,6)}$ to the $\beta$-function of $\lambda_{(2,8)}$. All these contributions are of two-loop order
and quantitatively not important to our analysis. First of all, the coefficients $\lambda_{(m,n)}$ are small and second, in the electroweak minimum these operators are suppressed by $\epsilon^m$ and hence do not contribute much to the scalar masses and VEVs. The sole exception to this argument is the operator $\bar\lambda\equiv\lambda_{(4,8)}$ that obtains a substantial $\beta$-function from the bottom quark, see (\ref{eq:contr_bottom}). It is also this operator, that produces a large $\beta$-function for $s \sim M$ and fosters the naive expectation that there might be a unstable direction in the scalar potential.

The EFT contains the same degrees of freedom as the UV theory but the two heavy FN quarks. The tree level scalar potential contains equation (\ref{eq:V0}) and the operator
\be
\frac{\bar\lambda}4 h^4 \left( \frac{s}{M} \right)^8 \, .
\ee
The one-loop contribution from the bottom quark induces a running of this operator. In turn, this operator also contributes to the one-loop effective potential through its contribution to the Higgs and flavon masses. This effect will be neglected since it is parametrically suppressed by powers of $w/M$ and $v/M$.

In the matching, the operator $\bar\lambda$ is matched to the results from the UV theory in (\ref{eq:heavy_stuff}). In essence, $\bar\lambda$ is a small positive number and basically vanishes at scale $M$~\footnote{
Notice that this matching also absorbs the divergences in the higher dimensional operators that are present in the EFT but are absent in the UV theory by construction.  
}. The five remaining parameters of the scalar potential are fixed by the scalar masses, VEVs and mixing as before. This is conceptually somewhat cumbersome since not all parameters are fixed at the same scale but numerically quite easy to implement. We define parameters at the scales depicted in figure \ref{figureEFTconstruction}.\\

\paragraph{The SM EFT (Region IV).} The next threshold occurs at the mass scale of the FN scalar mass and the running changes slightly for $h < m_{\sigma}$. We would like to setup again an EFT, but this time also integrate out the flavon degree of freedom. Below the threshold, we expand $m_{\sigma}$ in the VEVs in order to obtain a potential in which the logarithmic contributions are VEV-independent, just as in the contributions in equation (\ref{eq:heavy_stuff}). This is essential for the setup and it guarantees that, after renormalizing the operators of dimension four, the impact of the heavy FN flavon field is encoded in higher dimensional operators that cease importance at small energy scales. Contrary to the case of the FN quarks, the mass will not admit an expansion around $s\simeq 0$. Instead, it is natural to expand the mass around minimum of the potential~\footnote{Notice that when expanding the FN quark masses this difference is irrelevant and expanding around the minimum of the potential does in fact also lead to (\ref{eq:heavy_stuff}).}, $s \sim w$. This leads to $m_{\sigma}^2 \simeq 3\lambda_s s^2 - \lambda_s w^2 +\lambda_m h^2/2$ for small mixing and to
\bea
\frac{1}{4} m_{\sigma}^4 
\lp \log[m_{\sigma}^2/\mu^2] - \frac32 \rp &\to& \nn \\
&&\hskip -3 cm 
\left[ \frac94 \lambda_\sigma^2 L_s  + \frac{9 \lambda_m \lambda_s h^2}{16 w^2} - 
\frac{ 9 \lambda_m^2 h^4}{128 w^4} \right] (s^2 - w^2)^2 \nn \\
&&\hskip -3 cm + 
\left[ 3 w^2 \lambda_s^2 (L_s -1) + 
 \frac34 \lambda_m \lambda_s L_s h^2 + 
  \frac{3\lambda_m^2 h^4}{32 w^2} \right] (s^2 - w^2) \nn \\
&&\hskip -3 cm + 
\frac12 w^4 \lambda_s^2 (-3 + 2 L_s) + 
 \frac12 w^2 \lambda_m \lambda_s (-1 + L_s) h^2 \nn \\
&&\hskip -3 cm + 
 \frac{1}{16} \lambda_m^2 L_s h^4 \, ,
\eea
where $L_s = \log[2 w^2 \lambda_s/\mu^2]$. Again, from these renormalizable operators one can read off the contributions from the flavon to the various $\beta$-functions. This potential has to be matched to the low energy EFT (i.e.~the SM) that does not even contain the flavon as a dynamical degree of freedom. In order to integrate out the flavon, the equation of motion of the flavon has to be solved and reinserted into the effective action \cite{Burgess:2007pt}. This will induce further higher dimensional operators for the Higgs scalar that we will however neglect. The flavon direction in the SM is not present and thus we depict it as `grayed-out' region (Region IV) in figure~\ref{figureEFTconstruction}. The SM concludes the chain of EFTs that have to be set up to study the effective potential. In the next subsection we discuss some subtleties of the matching procedure before we present numerical results.

\subsection{Matching method and resummation \label{sec:Matching}}

In this section will discuss an issue in constructing the correct EFT in more detail: matching in the scalar sector and resummation. The issue is that we encounter a hierarchy problem in the UV theory. To be specific, the expansion in (\ref{eq:heavy_stuff}) contains contributions to the operator $h^2$ that are of order $M^2$. This leads to very large threshold effects. 

\begin{figure}[t]
		\centering
		\includegraphics[width=0.3\textwidth]{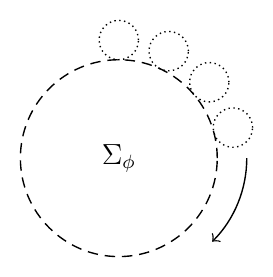}
	\caption{\small
		The Daisy diagrams that need to be resummed in the UV theory for accurate matching.}
	\label{fig:Daisies} 
\end{figure}
At first this does not seem to pose a problem. However, once the one-loop contributions of the Higgs and Goldstone bosons are inspected, a discrepancy in the two theories arises. The bare parameters are quite different in the two theories (due to the threshold effects) and hence the tree level masses in the scalar sector. While in the EFT the tree level Higgs mass is of EW scale, the tree level Higgs mass in the UV theory is of order $\kappa \, M^2$.

This discrepancy arises due to the breakdown of perturbation theory. The very same contribution that leads to a large threshold effect in the Higgs mass also leads to a large Higgs self-energy $\Sigma_\phi$. This large self-energy has to be resummed in order to recover the convergence of perturbation theory. The corresponding class of diagrams are the Daisy diagrams (see Fig.~\ref{fig:Daisies}) and the whole procedure is akin to the resummation performed in~\cite{Elias-Miro:2014pca, Martin:2014bca, Espinosa:2017aew} to solve the Goldstone boson catastrophe. 

In essence, this leads to an effective potential where the threshold effects also show up in the one-loop contributions of the scalar fields in the UV theory, e.g.
\bea 
\frac{n_G \kappa}{4}\left(-(m^2+\Delta m^2) +(\lambda+\Delta\lambda)\phi^2+\frac{\lambda_m}{2}w^2\right)^2\notag\\ 
&&\hskip -8cm \times\left[-\frac{3}{2}+\log\left(\frac{-(m^2+\Delta m^2) +(\lambda+\Delta\lambda)\phi^2+\frac{\lambda_m}{2}w^2}{\mu^2}\right)\right] \, .
\eea
The resulting effective potential can then be straight-forwardly matched to the IR theory since the arguments in the logarithm now 
coincide in the IR and UV theories. We do this numerically for all renormalizable operators in the EFT as well as the $\bar\lambda$ operator and at both threshold scales $\mu \sim m_s$ and $\mu \sim M$.

\subsection{EFT instabilities}\label{sec:Discussion1}

\begin{figure}[t]
	\begin{subfigure}[t]{0.31\textheight}
		\includegraphics[width=0.9\textwidth]{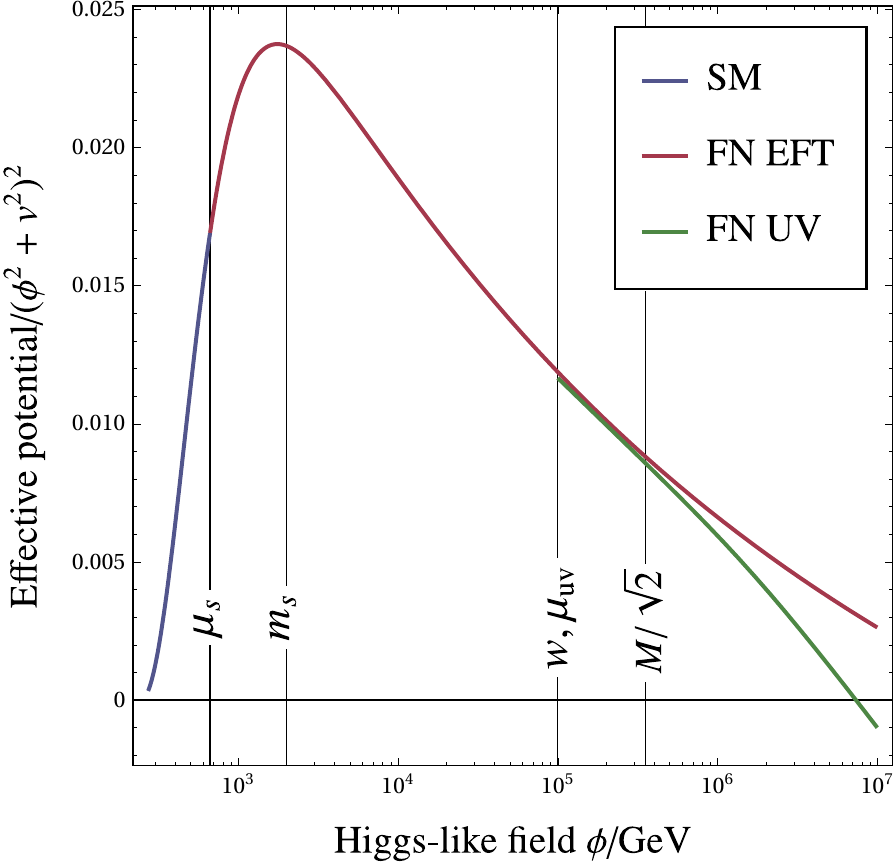}
	\end{subfigure}
	\begin{subfigure}[t]{0.31\textheight}
		\includegraphics[width=0.915\textwidth]{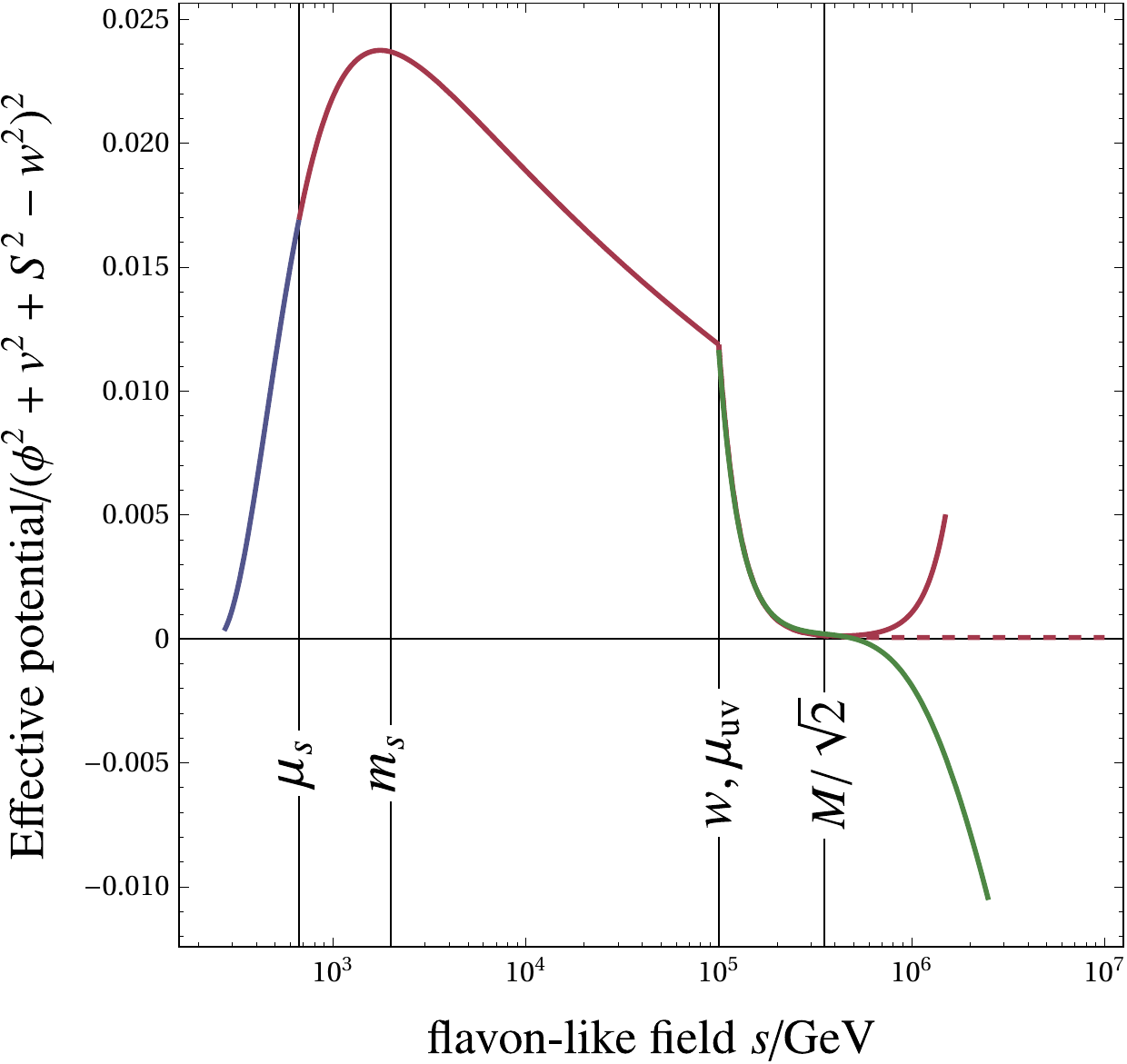}
	\end{subfigure}
	\caption{\small
Effective potential along the axes from figure~\ref{figureEFTconstruction}. 
The left part of the plots are identical and show the potential for fixed $s=w$ and variable $h$. Beyond this point, we show the potential along the $h$-direction (left) and $s$-direction with $h=w$ (right). The dashed line corresponds to the tree level EFT.
In this figure we have chosen the NP scales to be $m_{\sigma}=2\text{ TeV}$ and $M=5\times10^5 \text{ TeV}$. The measured Cabibbo angle then requires $w = \epsilon M \simeq 10^5 \text{ TeV}$. In both plots the scalar mixing angle is set to $\theta=0$.}
	\label{figureEFTtower1}
\end{figure} 

Consider first the effective potential in the regions below the FN threshold, in particular $w \ll (h \simeq s) \to M$ where instabilities from the presence of the $b$ quark are expected (Region II). To assess this regime, the theory is run from the electroweak scale $h=\mu \simeq v$ to the scale of the flavon VEV $h=\mu \simeq w$ and finally to the energy scale of the FN quarks along $h=s=\mu \to M$. All relevant logarithms are resummed and this is the regime where the instability in the effective potential should occur if the naive argument from the introduction was correct and the increased running in the quartic Higgs coupling was disastrous. From the analysis it should be clear by now that this will not be the case. In figure \ref{figureEFTtower1} we show that the flavon EFT including the light $b$ quark does not induce any instabilities below the Froggatt Nielsen scale in the regions $(h\rightarrow M,w)$ and $(w,s\rightarrow M)$, and that the instability only appears by matching the model dependent UV model we described in section~\ref{sec:FNEffectivePotential}.  
So in essence, for $w \gtrsim \epsilon M$ the running enhances the operator $\bar \lambda$ in the IR instead of destabilizing the potential in the UV.

However, other constraints can potentially be relevant. For example the operator $\bar \lambda$ could become very large due to running and violate perturbative unitarity in the limit where the mass $M$ is moved to very large scales rendering the EFT construction invalid. This is also not the case, since this would require a very large hierarchy between the UV scale $M$  and the flavon mass $m_s$ and in turn a tiny coupling $\lambda_s$. Thus, in general the $b$ quark has minimal impact on vacuum stability below the FN threshold.
 
\begin{figure}[t]
	\includegraphics[width=1.0\textwidth]{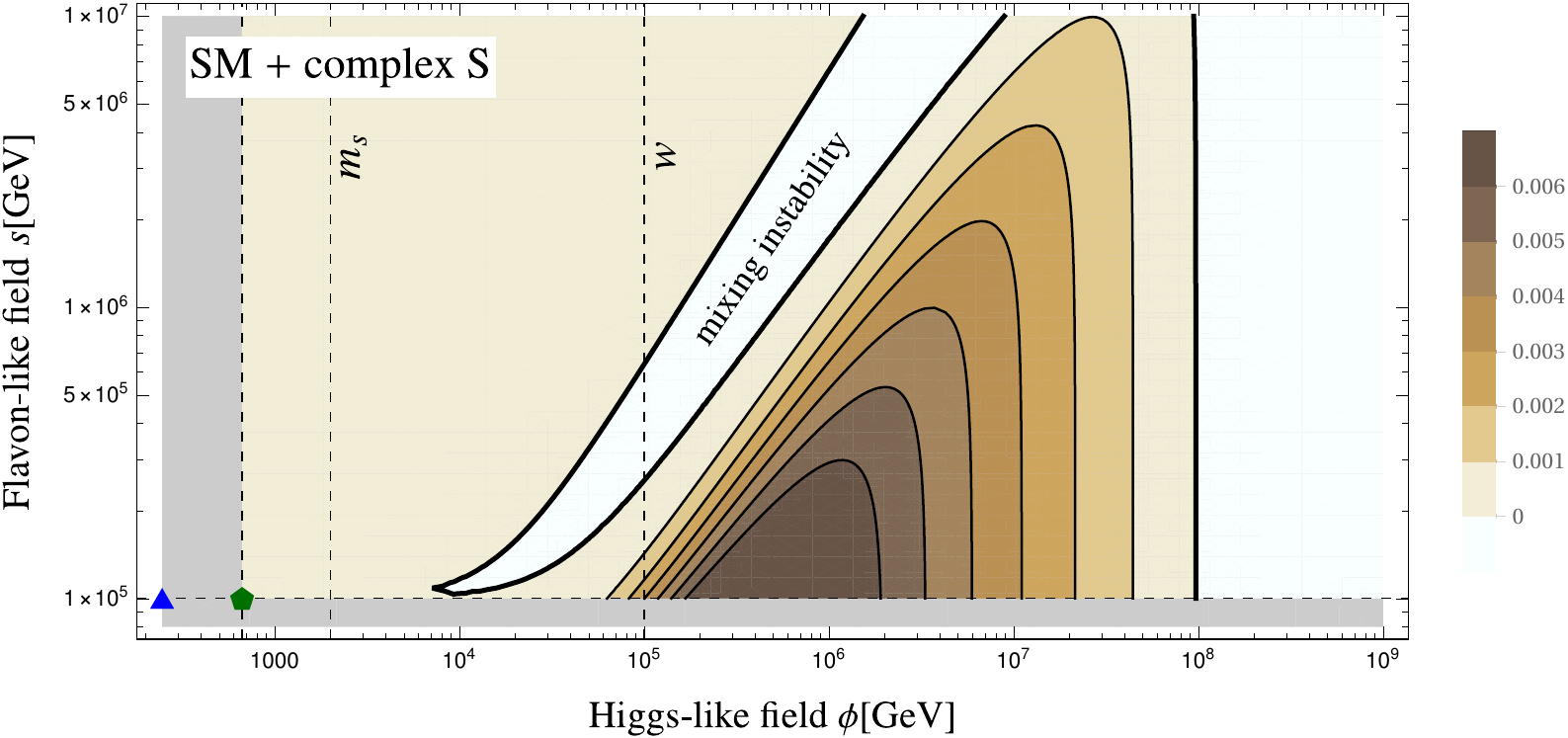}
	\caption{\small
		EFT landscape for the SM matched to a SM+complex scalar with \mbox{$m_\sigma=2\text{ TeV}$}, $w=10^5\text{ GeV}$ and $\theta=10^\circ$. The flavon direction is destabilized by mixing. The vacuum stability of the SM+complex scalar in a DM context has quantitatively been discussed in \cite{Gonderinger:2009jp,Gonderinger:2012rd}.}
	\label{figureEFTlandscape1}
\end{figure}

\begin{figure}[h]
	\includegraphics[width=1.0\textwidth]{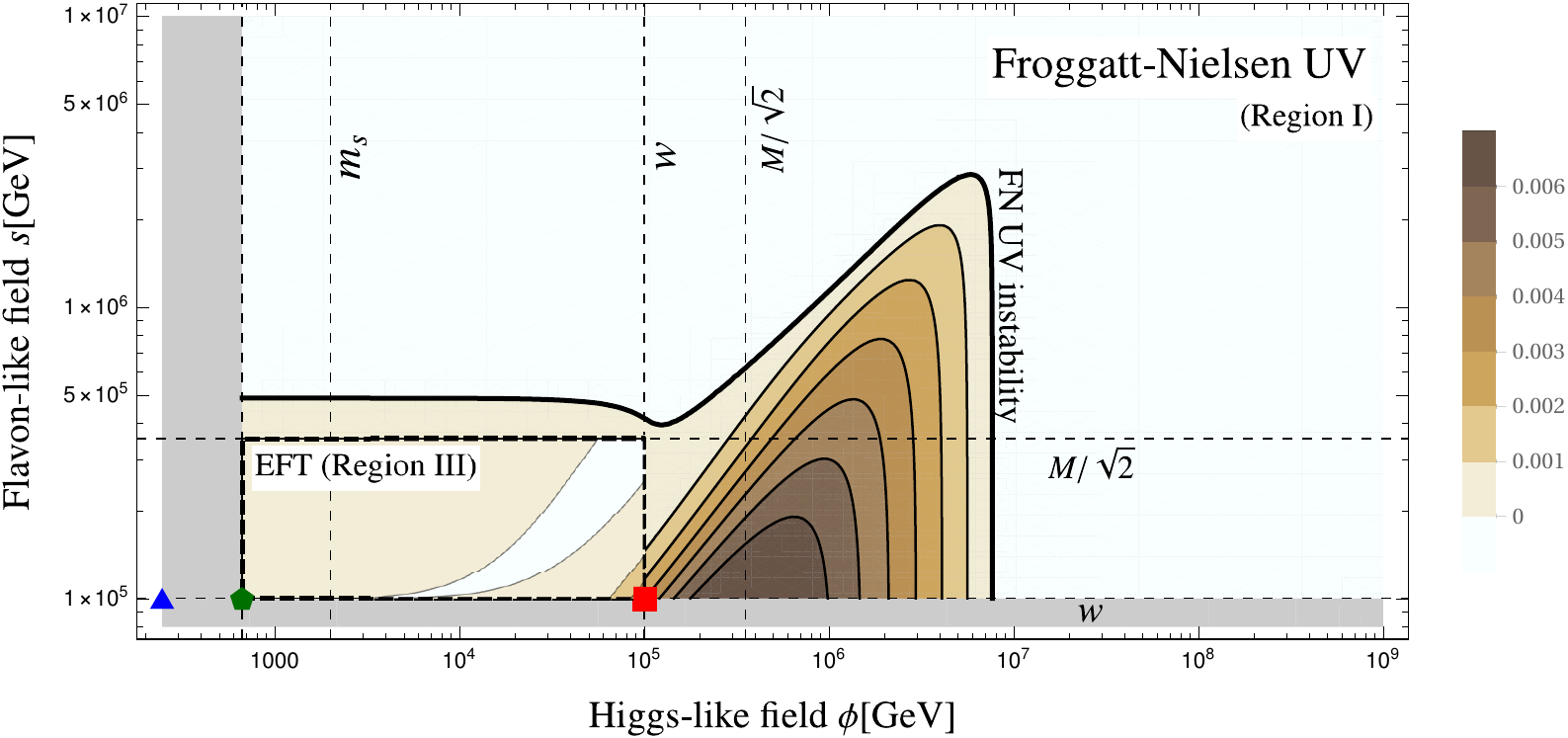}
	\caption{\small
		UV completion ($M=5\times10^5\text{ GeV}$, $m_\sigma=2\text{ TeV}$) with the UV model matched to the flavon EFT at $\mu=\phi=w$. The white island corresponds to the residual instability region from figure \ref{figureEFTlandscape1} that is left over once the light $b$ quark and higher dimensional operators are taken into account.}
	\label{figureEFTlandscape2}
\end{figure}

To visualize again the different effects, we first show the effective potential of the SM extended with a scalar but without the quarks (figure~\ref{figureEFTlandscape1}). 
We chose $\theta=10^\circ$ deliberately such that the mixing instability becomes severe. The point where we match the flavon theory to the SM EFT is indicated by a green pentagon whereas the EW minimum of the potential is indicated by a blue triangle. The instability develops in the would-be EFT region in a Froggatt-Nielsen model via tree level mixing.  In figure~\ref{figureEFTlandscape2}, we show in comparison the FN EFT matched to the UV at the red circle (the size of Region II thus shrinks to zero). The FN EFT does not develop an instability due to the $b$-quark contribution because of the presence of the higher dimensional operators at $s\sim M/\sqrt{2}$, in fact the higher dimensional operators even cure most of the unstable regions from figure~\ref{figureEFTlandscape1} to make the matching at the UV scale possible. In the bottom right it can be seen that the SM-like instability scale in the EFT is insensitive to displacements in $\langle s\rangle$ which refutes the naive expectation that the running in $\lambda$ from the FN contribution leads to an instability issue. The unstable island inside region III is due to the mixing operator which is a free parameter in the FN model.

Since the potential is in this regime to good approximation quadratic, the tunneling to this island happens via a Fubini bounce with action $S \simeq 8\pi^2/(3|\lambda|)$. Tunneling requires $\lambda<0.1$ which is never achieved in our model for the instability generated by mixing.

\FloatBarrier
\FloatBarrier

\section{Conclusion}

In this work, we studied the vacuum stability of Froggatt-Nielsen models. Main motivation for this analysis are models with varying Yukawa couplings, as employed in baryogenesis scenarios~\cite{Baldes:2016gaf}. In this context, the stability analysis leads to some open questions~\cite{Braconi:2018gxo} that partially are not well posed due to the fact that the cutoff of this setup is expected to be very low. Here, we demonstrate this issue by studying an explicit Froggatt-Nielsen setup, which is one possibility to induce varying Yukawa couplings (other examples are extradimensions or composite Higgs models~\cite{Bruggisser:2018mrt}). 

On the technical side, we use the effective potential improved by the renormalization group running. Above the UV threshold we use the full Froggatt-Nielsen model while below the threshold we use an effective theory that contains the SM (with a light b quark) and the flavon. 
It turns out that an accurate matching of the models also requires a resummation of self-energies (see Section~\ref{sec:Matching}).

A new source of instability can be mixing between the Higgs and flavon, but taking phenomenological constraints into account leads to a lifetime of the electroweak vacuum that exceeds the age of the Universe. 

Our main conclusion is that the problems of stability are generally not more severe than in the Standard Model. Universally, we find that instead of leading to stability issues in the UV, the FN dynamics is such that the effective potential is increased once the Yukawa couplings start to vary. For example, if the Yukawa couplings change due to a phase transition in the FN sector, the potential difference between these two phases is so large that it overcompensates the increase in running in the Higgs quartic coupling that naively leads to the instability. This is essentially due to the fact that the operators that contribute to the increased running in the Higgs quartic coupling (due to a dependence on the flavon VEV) are generated by additional FN quarks and hence these operators vanish close to the UV scale. The behavior in the far UV depends crucially on the concrete model which is why no definite conclusions can be drawn in this regime. Our analysis can be straight-forwardly generalized to a setup where all fermions obtain masses via the FN mechanism. 

\section*{Acknowledgments}

We would like to thank Iason Baldes for discussions on the phenomenology of the FN setup.
TK is supported by the Deutsche Forschungsgemeinschaft under Germany's Excellence Strategy – EXC 2121 ``Quantum Universe'' – 390833306.

\newpage
\appendix
\section{Comparing with experimental constraints}\label{sec:ExperimentalBounds}
In this section we discuss various experimental constraints that are of relevance in a realistic Froggatt-Nielsen scenario. We discuss the two kind of constraints which are the most relevant to our discussion. First we briefly discuss bounds from flavor physics which put the most stringent constraints in realistic Froggatt-Nielsen scenarios. We discuss constraints for the scalar mixing angle in order to get an estimate on what values are reasonable to consider in the main part of our analysis. 

\subsection{Bounds from meson mixing}\label{sec:BoundsFromMesonMixing}
The dynamics of Meson systems are expected to yield strong experimental constraints on a Froggatt-Nielsen mechanism realized at $\text{TeV}$ scale. Meson bounds for the Froggatt-Nielsen mechanism have already been discussed in full detail in \cite{Bauer:2016rxs}. The Lagrangian of the SM Yukawa couplings is given by
\begin{equation}
\lag_{\text{yuk}}=-\left(y_{ij}^u\ \barquarkL{q}{i}\Higgs\quarkR{d}{j}+y_{ij}^d\ \barquarkL{q}{i}\tilde\Higgs\quarkR{u}{j}\right)+h.c\ .
\end{equation}
In the effective FN EFT this becomes
\begin{equation}
\lag_{\text{yuk}}=-\left[Y_{ij}^u\left(\frac{\sqrt{2}S}{M}\right)^{n^d_{ij}}\ \barquarkL{q}{i}\Higgs\quarkR{d}{j}+Y_{ij}^d\left(\frac{\sqrt{2}S}{M}\right)^{n^u_{ij}}\ \barquarkL{q}{i}\tilde\Higgs\quarkR{u}{j}\right]+h.c\label{EqnBrokenYukawaLagrangian}
\end{equation}
with $Y_{ij}^{d,u}\sim\orderone$. In the broken phase we have
\begin{equation}
H=\frac{1}{\sqrt{2}}\begin{pmatrix}
0 \\ v+\phi(x)
\end{pmatrix}\qquad\text{and}\qquad S=\frac{w+\sigma(x)+ia(x)}{\sqrt{2}}.
\end{equation} 
We fit the $Y_{ij}^{u,d}$ parameters to obtain measured observables. Choosing 
\begin{align}
Y^d&=\begin{pmatrix}
0.51 - 0.18i && -0.38 + 1.11i && -1.02 -0.58i\\
-1.33-0.43i && -0.77-0.68i && -0.38+0.99i \\
-0.94-0.88i && -0.04+0.52i && 0.55+0.82i
\end{pmatrix},\qquad\\
Y^u&=\begin{pmatrix}
-0.63 - 0i&&  -1.38 + 0.05 i && -1.41-0.25 i \\
0.09+0.52 i && -0.26+0.55 i && 0.55 + 0.96 i \\
0.73+0.24 i && 0.56 -0.16 i && 0.47 -0.21 i
\end{pmatrix},
\end{align}
gives acceptable values matching the observed quark masses and mixing angles. Here the suppression of the SM masses and CKM matrix elements is set by the Cabibbo angle \mbox{$\theta_{12}\sim\epsilon\sim\frac{w}{M}=0.2$} and the Froggatt-Nielsen charges $FN(q)$. We use the following convention
\begin{align}
n_{ij}^d&=FN(\quarkL{q}{i})+FN(\quarkR{d}{j})-FN(\Higgs),\\
n_{ij}^u&=FN(\quarkL{q}{i})+FN(\quarkR{u}{j})+FN(\Higgs).
\end{align}
and the following charge assignment\\
\begin{equation}
FN\begin{pmatrix}
\quarkL{q}{u,c,t}\\
\quarkR{q}{u,c,t}\\
\quarkR{q}{d,s,b}\\
\end{pmatrix}=\begin{pmatrix}
3&2&0\\5&1&0\\4&2&2
\end{pmatrix},\qquad FN(\Higgs)=0, \qquad FN(\Flavon)=-1.
\end{equation}
Expanding (\ref{EqnBrokenYukawaLagrangian}) to first order in $v$  gives
\begin{equation}
-\lag_{\text{yuk}}^{\text{broken}}=\frac{v}{\sqrt{2}}y_{ij}\epsilon^{n_{ij}}\barquarkL{q}{i}\left(1+\frac{\phi}{v}+\frac{n_{ij}\ \sigma+in_{ij}\ a}{w}\right)\quarkR{q}{j} \, .
\end{equation}
\begin{table}[t!]
\begin{center}
\begin{tabular}{llllll}
		&                                          & \multicolumn{1}{c}{Kaon $C_K$}            & \multicolumn{1}{c}{D-Meson $C_D$} & \multicolumn{1}{c}{B-Meson $C_{B_d}$}     & \multicolumn{1}{c}{B-Meson $C_{B_s}$}     \\ \hline \hline
		\multicolumn{1}{c}{(2008)} & $\left|\text{Re}\ C_2\right|$            & \multicolumn{1}{c}{$1.9\times10^{(-14)}$} &                                   &                                           &                                           \\
		& $\left|\text{Im}\ C_2\right|$            & $9.3\times10^{(-17)}$                     &                                   &                                           &                                           \\
		& $\left|\text{Re}\ C_4\right|$            & $3.6\times10^{(-14)}$                     &                                   &                                           &                                           \\
		& $\left|\text{Im}\ C_4\right|$            & $1.9\times10^{(-17)}$                     &                                   &                                           &                                           \\
		& \multicolumn{1}{c}{$\left|C_2\right|$}   &                                           & $1.6\times10^{(-13)}$             & \multicolumn{1}{c}{$7.2\times10^{(-13)}$} & \multicolumn{1}{c}{$5.6\times10^{(-11)}$} \\
		& \multicolumn{1}{c}{$\left|C_4\right|$} &                                           & $4.8\times10^{(-14)}$             & \multicolumn{1}{c}{$2.1\times10^{(-13)}$} & \multicolumn{1}{c}{$1.6\times10^{(-11)}$} \\ \hline \hline
		\multicolumn{1}{c}{(2014)} & $\left|\text{Re}\ C_2\right|$            & \multicolumn{1}{c}{$5.2\times10^{(-15)}$} & $1.6\times10^{(-13)}$                                   &                                           &                                           \\
		& $\left|\text{Im}\ C_2\right|$            & $1.8\times10^{(-17)}$                     & $2.3\times10^{(-14)}$                                    &                                           &                                           \\
		& $\left|\text{Re}\ C_4\right|$            & $1.0\times10^{(-15)}$                     & $4.2\times10^{(-14)}$                                    &                                           &                                           \\
		& $\left|\text{Im}\ C_4\right|$            & $3.7\times10^{(-18)}$                     & $6.8\times10^{(-15)}$                                   &                                           &                                           \\
		& \multicolumn{1}{c}{$\left|C_2\right|$}   &                                           &             & \multicolumn{1}{c}{$2.8\times10^{(-13)}$} & \multicolumn{1}{c}{$3.8\times10^{(-12)}$} \\
		& \multicolumn{1}{c}{$\left|C_4\right|$} &                                           &              & \multicolumn{1}{c}{$8.4\times10^{(-14)}$} & \multicolumn{1}{c}{$1.2\times10^{(-12)}$} \\ \hline \hline
\end{tabular}
\caption{\label{table_WilsonCoefficients}\small List of used values for the Wilson coefficients (in units $1/\text{GeV}^2$) \cite{Bona:2007vi,Bevan:2014cya}. We only use absolute values of the Wilson coefficients. For the Kaon system generalized bounds exist depending on the sign of the contribution which would slightly improve the bound.}
\end{center}
\end{table}
We see that the FN mechanism introduces flavor changing neutral currents (FCNCs) at $1/{m_\sigma^2}$ level but with suppressed couplings proportional to $\epsilon^{n_{ij}}$, namely
\begin{equation}
	g^{u,d}_{ij}=Y^{u,d}_{ij}n_{ij}\frac{v}{\sqrt{2}w}\epsilon^{n_{ij}^{u,d}}.
\end{equation}
The FCNC contribution to meson mixing can be parametrized by the $\Delta F=2$ operators
\begin{equation}
H_{\text{eff}}^{\Delta F=2}=\sum_{k=1}^{5}C_k^{ij}Q_k^{ij}+\sum_{k=1}^{3}\tilde{C}_k^{ij}\tilde{Q}_k^{ij},
\end{equation}
where for our model the only contributing operators are 
\begin{equation}
Q_2=\left(\barquarkR{q}{i}\quarkL{q}{j}\right)^2,\qquad\tilde{Q}_2=\left(\barquarkL{q}{i}\quarkR{q}{j}\right)^2,\qquad Q_4=\left(\barquarkR{q}{i}\quarkL{q}{j}\right)\left(\barquarkL{q}{i}\quarkR{q}{j}\right).
\end{equation}
\begin{figure}[t!]
	\begin{subfigure}[h]{0.5\textwidth}
		\includegraphics[width=0.92\textwidth]{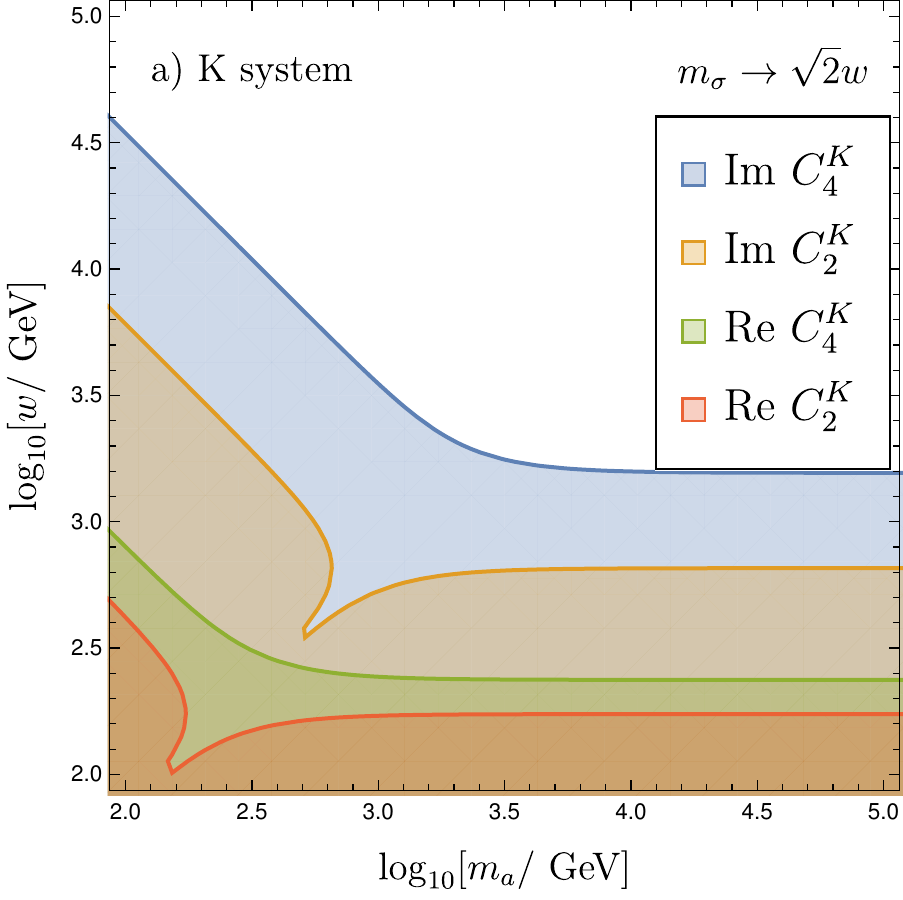}
		\label{figureKaonMixingBounds1}\\
		\includegraphics[width=0.92\textwidth]{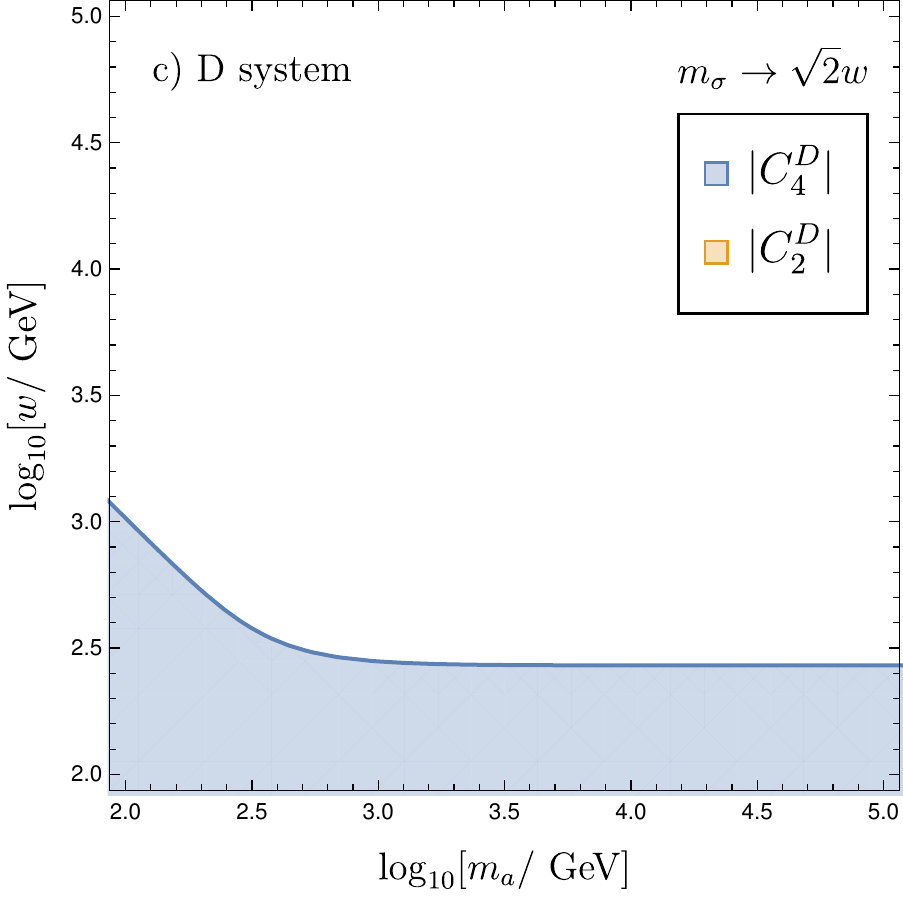}
		\label{figureDMesonMixingBounds1}\\
		\includegraphics[width=0.92\textwidth]{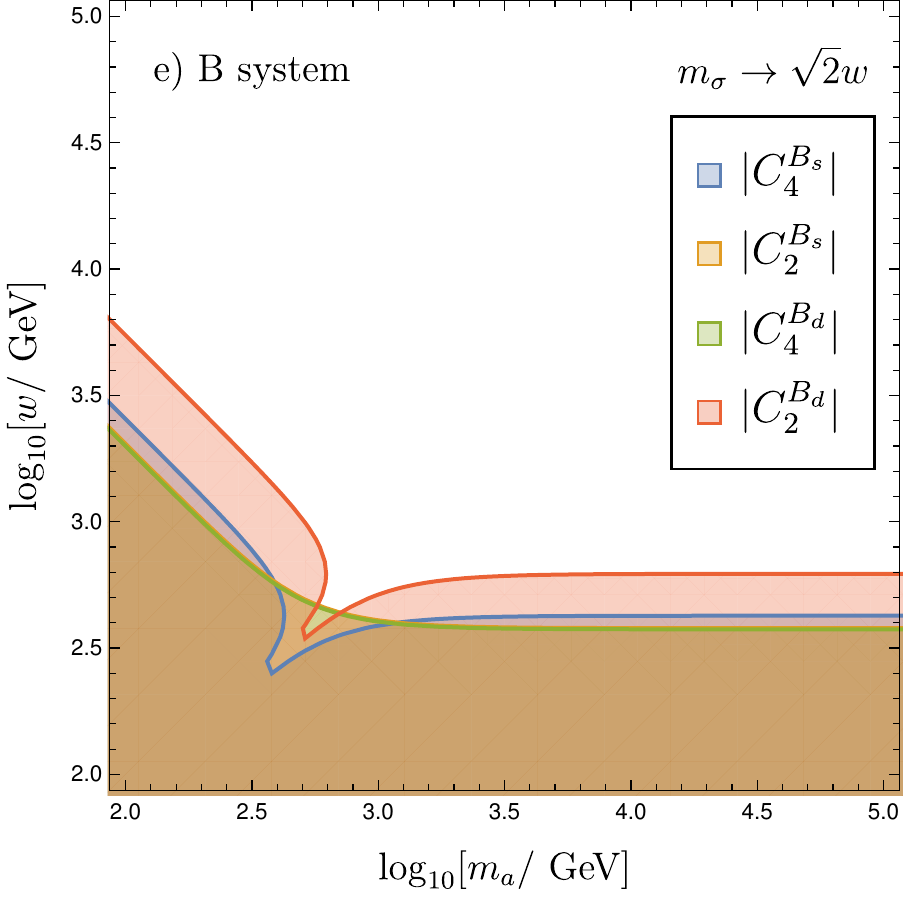}
		\label{figureBMesonMixingBounds1}	
	\end{subfigure}
	\begin{subfigure}[h]{0.5\textwidth}
		\includegraphics[width=0.92\textwidth]{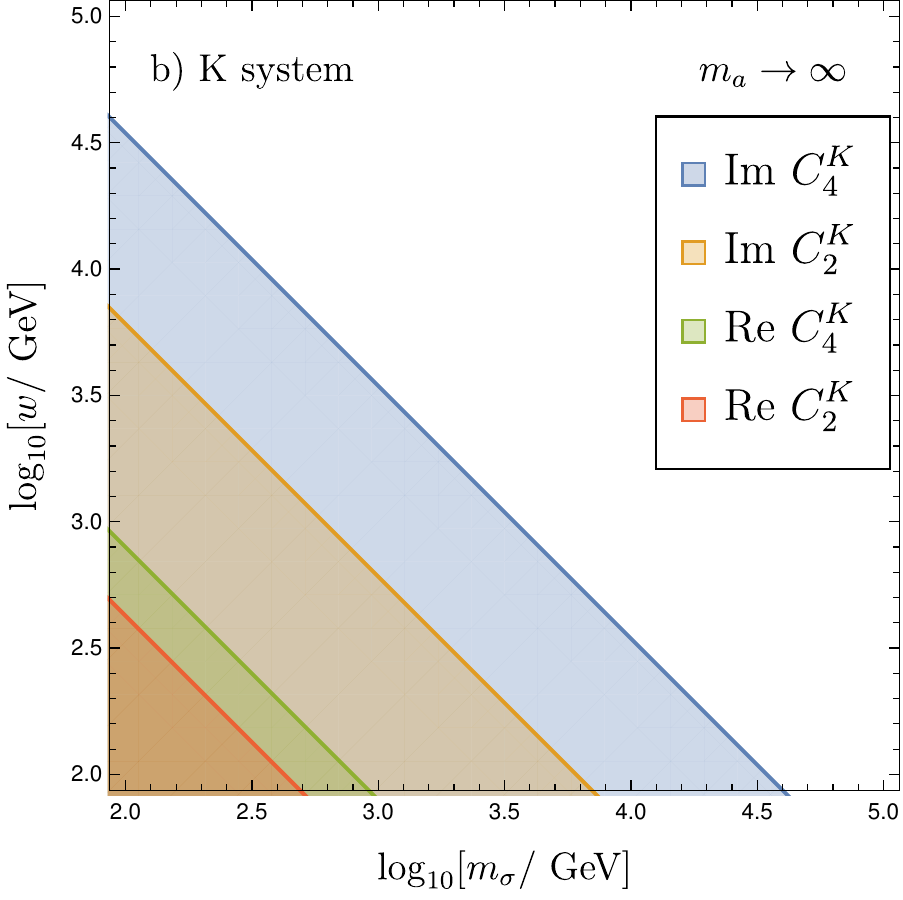}
		\label{figureKaonMixingBounds2}\\
		\includegraphics[width=0.92\textwidth]{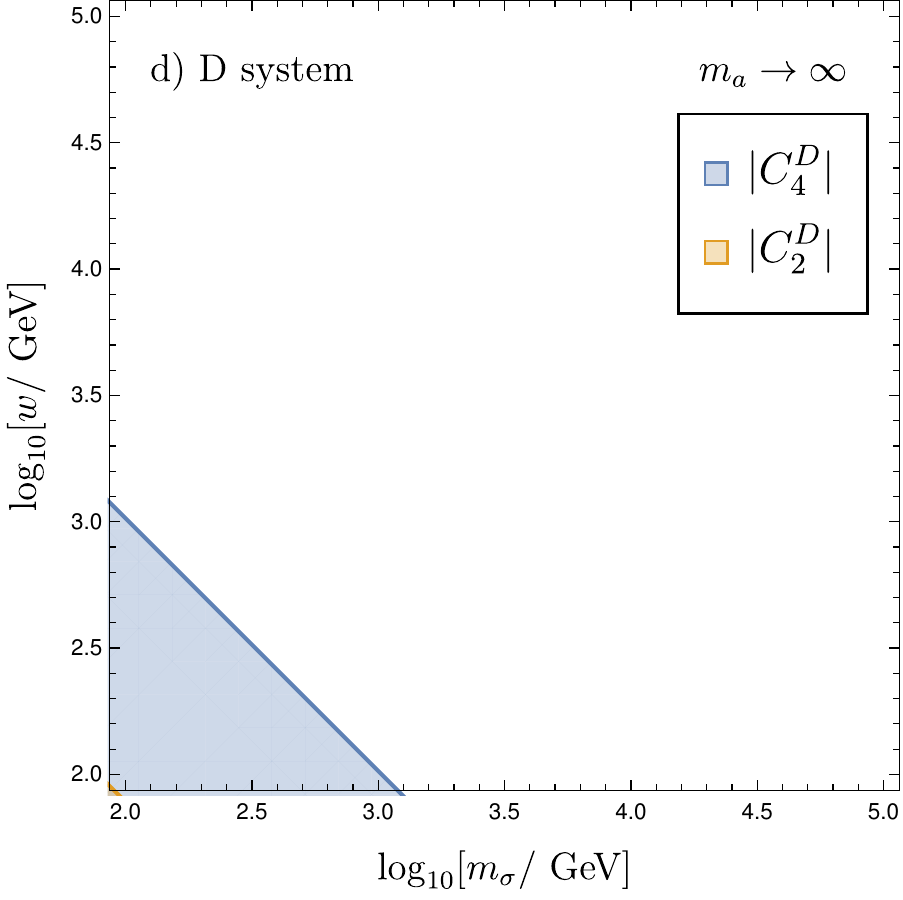}
		\label{figureDMesonMixingBounds2}\\
		\includegraphics[width=0.92\textwidth]{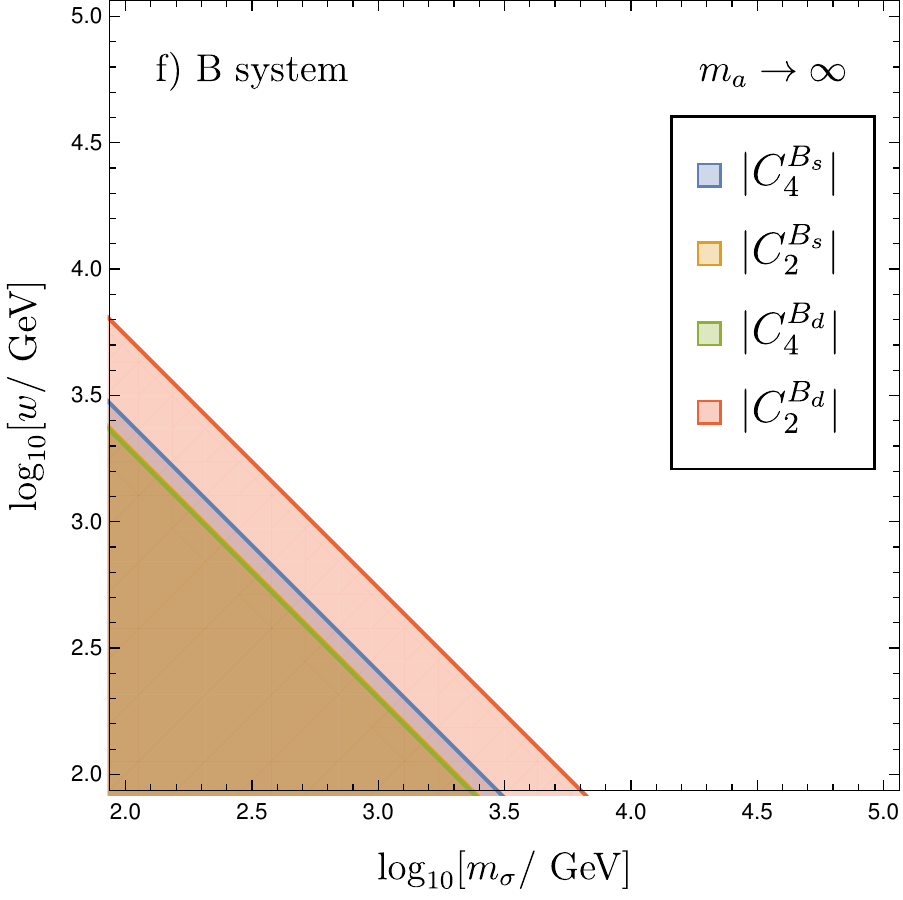}
		\label{figureBMesonMixingBounds2}	
	\end{subfigure}
	\caption{\small
		Meson mixing exclusion limits for all K, D, $\text{B}_{\text s}$, $\text{B}_{\text d}$ systems in different regions of the parameter space (left: $m_\sigma=\sqrt{2}w$, right: $m_a\rightarrow\infty$)}
	\label{figureMesonMixing}
\end{figure}The corresponding tree level Wilson coefficients $C_2$ from t-channel scalar scattering and $C_4$ from s-channel scalar scattering are 
\begin{eqnarray}
&C_2=\left(g^*_{ji}\right)^2\left(\frac{1}{m_\sigma^2}-\frac{1}{m_a^2}\right) \;,\qquad\tilde{C}_2=\left(g_{ij}\right)^2\left(\frac{1}{m_\sigma^2}-\frac{1}{m_a^2}\right)\; ,\;\notag\\
&\hspace{0.5cm}C_4=\frac{g_{ij}g_{ji}}{2}\left(\frac{1}{m_\sigma^2}+\frac{1}{m_a^2}\right).\;\label{equationWilsonCoefficients} 
\end{eqnarray}

The numerical values of these are given in table~\ref{table_WilsonCoefficients} from which we will use the 2014 values. We can use~(\ref{equationWilsonCoefficients}) to directly set constraints on the \mbox{(pseudo-scalar) flavon} masses and on the flavon VEV. We depict these bounds in figure \ref{figureMesonMixing}. The strength of the bound strongly depends on both, the experimental bound on $C_i$ and the suppression factor $n_{ij}g_{ij}$ in the coupling matrices. The most stringent bound arises from the Kaon-System, whereas the D meson system is the least sensitive. Note that the factor $g_{ij}$ allows for some freedom because of different possible choices of FN charges. For instance the Kaon bound would be strengthened by choosing $FN(d_R)=3$ instead of $FN(d_R)=4$. \\

\FloatBarrier
\subsection{Bounds from scalar mixing}
The scalar mixing angle $\theta$ between the Higgs and the flavon is constrained by several techniques (see \cite{Robens:2015gla,Efrati:2014uta} for an overview). Early constraints for $\theta$ come from the Z-pole measurement at LEP  \cite{Barate:2003sz} which constrains the signal strength and thus $g_{HZZ}\sim\cos\theta$,
\begin{equation}
\xi^2=\left(\frac{g^{BSM}_{HZZ}}{g^{SM}_{HZZ}}\right)^2\text{BR}\left(H\rightarrow \text{SM}\right),
\end{equation}
but only for scalars $H$ with mass below $114$ GeV. After the Higgs discovery a general bound on the mixing angle can be obtained from a combined fit of all Higgs signal channels from LHC measurements with the use of \texttt{HiggsSignals} \cite{Bechtle:2013xfa} including all Higgs signal data from LHC Run-1 + \mbox{Run-2$\left(36\text{ fb}^{-1}\right)$} from Atlas and CMS. The general bound for a scalar with branching ratio $\text{BR($H\rightarrow $\text{ SM})}=1$ is
\begin{align}
\theta< \begin{cases}
\sim 10 \text{ degree} & m_\sigma \lesssim 114 \text{ GeV}\text{, LEP bound}  \\
16.8 \text{ degree} & m_\sigma \sim 1 \text{ TeV}\text{, LHC bound} 
\end{cases}.
\end{align}
We compare these constraints with electroweak precision physics as parametrized by the oblique (or Peskin-Takeuchi) parameters $S$, $T$, $U$ \cite{Peskin:1992ds}, with the current best fit values~\cite{Baak:2012kk}
\begin{equation}
	S=0.05\pm 0.11,\qquad T=0.09\pm 0.13,\qquad U=0.01\pm 0.11.
\end{equation}
To leading log approximation and $\mu=m_s$ the oblique parameters are
\begin{equation}
S=
-\frac{\pi}{6}\left(1-\cos\theta^2\right)\log\frac{m_\phi}{m_\sigma},\quad T=\frac{3 m_Z^2}{8 \pi m_W^2}\left(1-\cos\theta^2\right)\log\frac{ m_\phi}{m_\sigma},\quad U=0.
\end{equation}
By performing a $\chi^2$ test, one sees that the EWPO bound falls short to the LHC bound.  

\begin{figure}[t!]
	\begin{subfigure}[h]{0.47\textwidth}
		\includegraphics[width=1.0\textwidth]{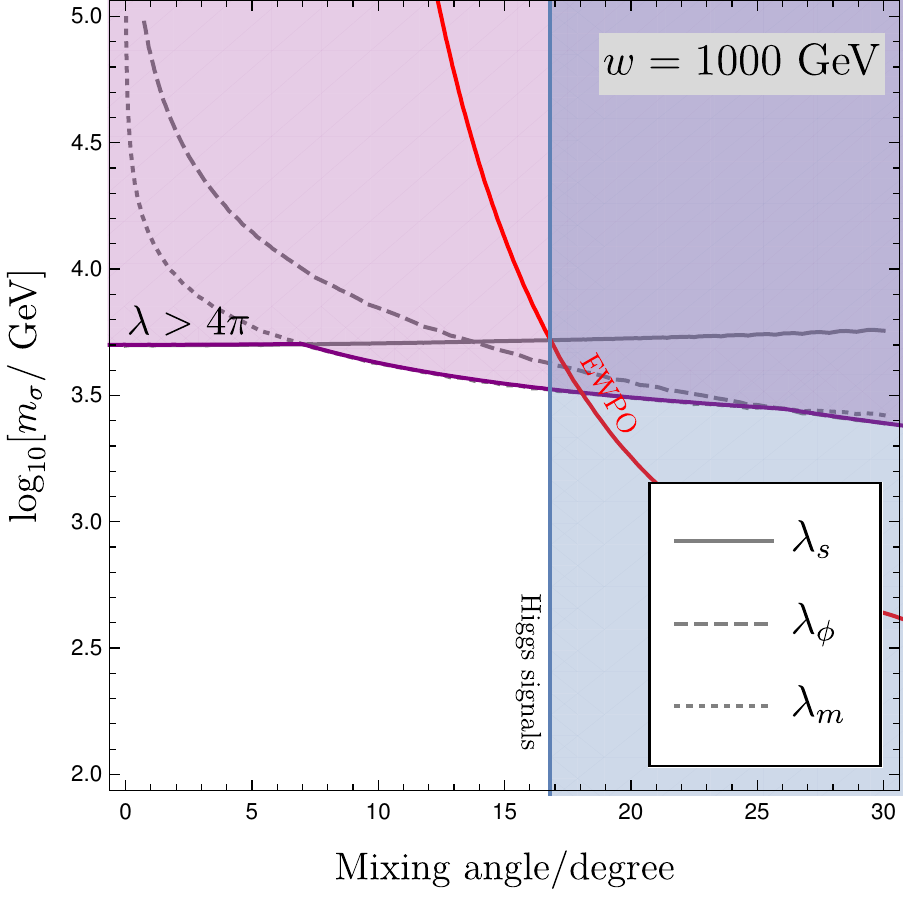}
		
	\end{subfigure}
	\begin{subfigure}[h]{0.47\textwidth}
		\includegraphics[width=1.0\textwidth]{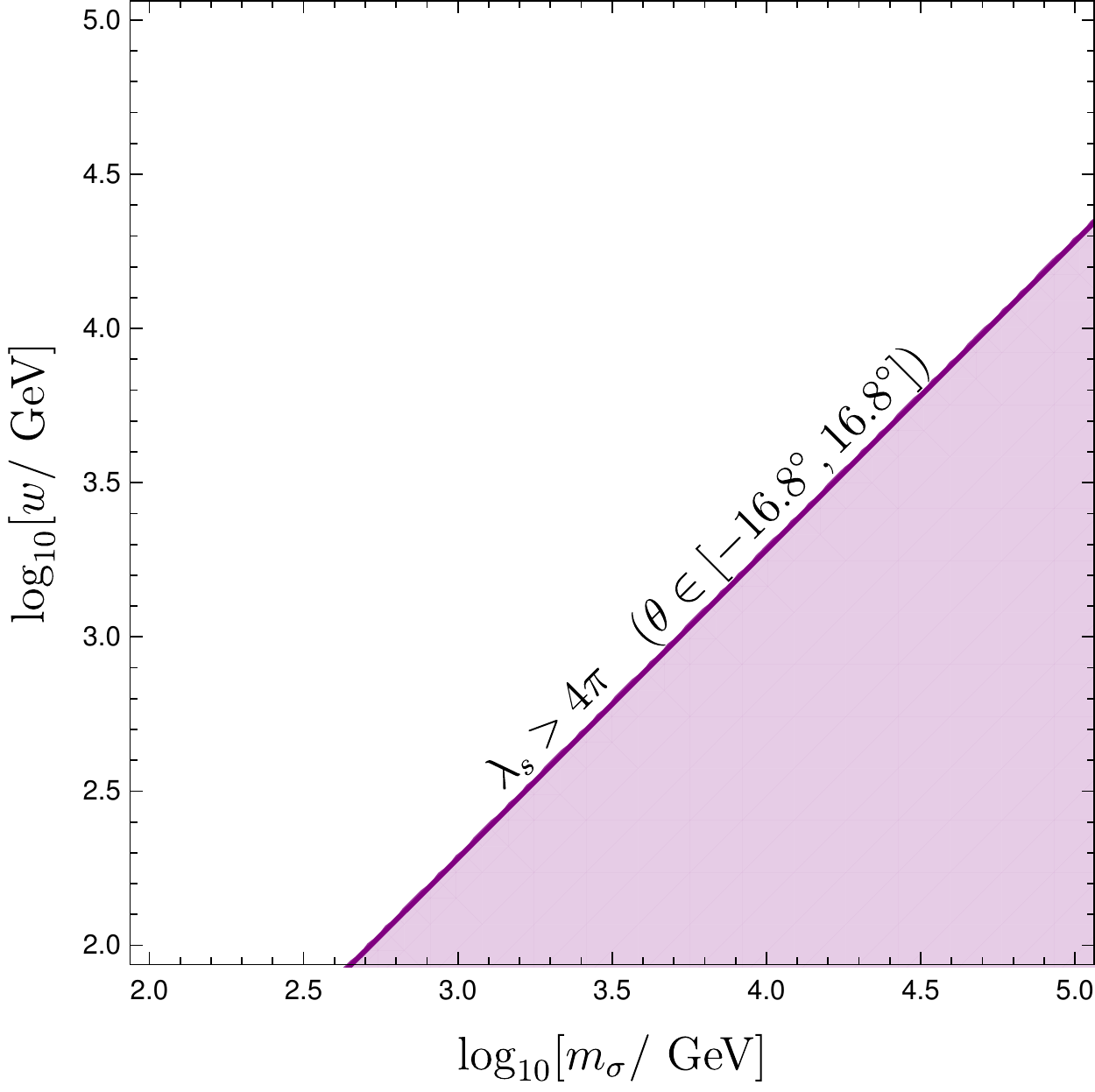}
	\end{subfigure}
	\caption{\small
On the left the perturbativity bound for varying flavon mass and varying scalar mixing angle is shown. The flavon VEV is set to $w=1000\text{ GeV}$. On the right the perturbativity constraint is shown varying flavon mass and flavon VEV.}\label{figurePerturbativitybound1}
\end{figure}

For larger flavon masses than TeV scale the mixing angle is suppressed by the flavon VEV and thus the scalar mixing becomes automatically small and hard to detect unless $\lambda_m$ becomes large because of
\begin{equation}
\tan 2\theta=\frac{\lambda_{m}v w}{\lambda_h v^2-\lambda_s w^2}\sim\lambda_m \frac{v}{w}.
\end{equation}
Consequently, this relation can also be read as a theoretical constraint on the mixing angle by requiring perturbativity. Requiring $\lambda_i < 4\pi$ for $\lambda_{m},\lambda_{h},\lambda_{s}$ thus also limits the allowed mixing. From this 
\begin{align}
2\theta\lesssim \begin{cases}
10 \text{ degree} & m_\sigma \gtrsim 1 \text{ TeV}, w\sim \order{\text{TeV}}\\
\text{unconstrained} & m_\sigma \gtrsim 1 \text{ TeV}, w\gg \order{\text{TeV}}\\
\end{cases}
\end{align}  
Information on the exact value of $w$ could be extracted in principle from a precision measurement of the cubic coupling $\lambda_{\phi\phi\phi}$ which is experimentally unfeasible at the moment. We show all bounds in figure~\ref{figurePerturbativitybound1}.

\FloatBarrier

\section{Running and beta functions}\label{appendixBetaFunctions}

Here we give a short summary of the RG-running in the EFT region of FN that we considered. We further give the $\overline{MS}$ parameters used throughout this paper as well as the used beta functions. For the SM, the resummation of large logs in the effective potential by the RG group is well understood, in particular the running of the top Yukawa itself leads to a higher instability scale $\Lambda_I$ already at 1-loop. At higher loop, threshold corrections push $\Lambda_I$ even further to the region where the SM is metastable. As input parameters for the SM we have chosen\footnote{Throughout the paper we use a slightly lighter top quark mass $y_t=0.95$ instead of $y_t=1$. We do this in order to recover the known instability scale from three-loop calculations that lead to additional threshold effects not present at 1-loop.} 
\begin{equation}
\begin{tabular}{llllllll}
\multicolumn{1}{l|}{$g(v)$} & \multicolumn{1}{l|}{$g'(v)$} & \multicolumn{1}{l|}{$g_3(v)$}  & \multicolumn{1}{l|}{$y_t(v)$}  & \multicolumn{1}{l|}{$y_b(v)$}  & \multicolumn{1}{l|}{$m^2(v)$}  & \multicolumn{1}{l}{$\lambda(v)$}  \\ \hline
\multicolumn{1}{l|}{0.654} & \multicolumn{1}{l|}{0.350} & \multicolumn{1}{l|}{1.128} & \multicolumn{1}{l|}{0.95} & \multicolumn{1}{l|}{0.02} & \multicolumn{1}{l|}{7812\text{ GeV}} & 0.12909 \\
&                        &                         &                         &                         &                         &   
\end{tabular}\ .
\end{equation}
The beta functions are defined by
\be
\beta^{(\text{full})}=\sum_i \kappa^i \beta^{(i)}
\ee
Here we give the beta functions of the Standard model with an additional Froggatt-Nielsen mechanism for the generation of the b quark mass term. We neglect all lighter than b quarks as well as leptons which means that we also do not discuss the running of SM mixing angles. We work in the $\overline{MS}$ scheme which means that we account for different particle thresholds by hand. We differentiate between the mass scale of the new flavon $m_s$ and in principle also the mass scale of the heavy fermion sector $M/\sqrt{2}$. The running in the UV theory depends however on the UV completion which is why we consider the running only in the region $\mu<M/\sqrt{2}$.
The anomalous dimensions are
\begin{align}
\gamma^{(1)}_\phi&=-3y_t^2-3y_b^2+\frac{3}{4}(3g^2+g'^2),\\
\gamma^{(1)}_s&=0.
\end{align}
 Using the Heaviside theta $\theta_m=\theta\left(\mu-m\right)$: 
\begin{align}
\beta^{(1)}_\Lambda&=2m_h^4+\theta_m\frac{m_s^4}{2},\\
\gamma^{(1)}_{m_h^2}&=12\lambda_h+\theta_m\frac{m_s^2}{m_h^2}\lambda_{m},\\
\gamma^{(1)}_{m^2_s}&=\lambda_s+\theta_m\frac{m^2}{m_s^2}\lambda^2_{m},
\end{align}
\begin{align}
\beta_{\lambda}^{(1)}&=\left(24\lambda_h^2-6y_t^4-6y_b^4+\frac{9}{8}g^4+\frac{3}{4}g^4 g'^2+\frac{3}{8}g'^2\right)+\theta_m\frac{\lambda_{m}^2}{2}-\notag\\&\qquad-
\underbrace{\left(+3\lambda_h g'^2+9\lambda_h g^2-12\lambda_h y_t^2-12y_b^2\lambda_h\right)}
_{=4\lambda\gamma^{(1)}_\phi}\label{quarticbetafunctionSM},\\
\beta^{(1)}_{\lambda_s}&=\theta_m\left(18{\lambda_s}^2+\frac{\lambda_{m}^2}{2}-4\gamma^{(1)}_s\lambda_{s}\right),\\
\beta^{(1)}_{\lambda_{m}}&=\theta_m\left(\lambda_m\left(6\lambda_h+4\lambda_{m}+6\lambda_s\right)-2\left(\gamma^{(1)}+\gamma_s^{(1)}\right)\lambda_{m}\right),\\
\beta_{g_s}^{(1)}&=-7 g_s^3,\\
\beta_{g}^{(1)}&=-\frac{19}{6} g^3,\\
\beta_{g'}^{(1)}&=+\frac{41}{6} g'^3,\\
\beta_{y_t}^{(1)}&=y_t\left(\frac{3}{2}y_t^2-\frac{3}{2}y_b^2+3y_t^2+3y_b^2-8g_s^2-\frac{9}{4}g^2-\frac{17}{12}g'^2\right),\\
\beta_{y_b}^{(1)}&=0.
\end{align}
In contrast to the SM the bottom quark Yukawa is not fundamental in the FN EFT which is why here $y_b=0$ for all $\mu$. For completeness the bottom quark beta function in the SM is given by
\begin{equation} \beta_{y_b}^{(1)}=y_b\left(\frac{3}{2}y_b^2-\frac{3}{2}y_t^2+3y_b^2+3y_t^2-8g_s^2-\frac{9}{4}g^2-\frac{3}{20}g'^2\right).
\end{equation}
The bottom quark in FN rather contributes to the presence of an effective operator with coupling $\bar\lambda$ which itself runs. The leading contribution is
\begin{equation}
\beta_{\bar{\lambda}}^{(1)}=-\frac{n_b}{2} Y_b^4 -(\gamma_\phi^{(1)}+2\gamma_s^{(1)})\bar\lambda.
\end{equation} 
Note that there are contributions to the beta functions due to other higher dimensional operators, e.g.

\begin{equation}
\includegraphics[width=0.3\textwidth]{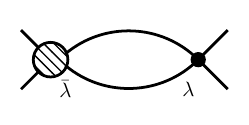},
\end{equation}

but after all these are higher-loop order effects which is why we neglect them.

\FloatBarrier
\bibliography{bib}
\bibliographystyle{JHEP}

\end{document}